\documentclass[a4paper,preprintnumbers,amsmath,amssymb]{revtex4}

\usepackage{graphicx}
\usepackage{dcolumn}
\usepackage{bm}
\usepackage{epstopdf}

\newcount\driver
\newcount\bozza

\font\ottorm=cmr8 scaled\magstep1 
scaled\magstep1                  
scaled\magstep1 \font\msytw=msbm10 scaled\magstep1
 
scaled\magstep1                 \font\indbf=cmbx10 scaled\magstep2

scaled \magstep2

{\count255=\time\divide\count255 by 60
\xdef\hourmin{\number\count255}
        \multiply\count255 by-60\advance\count255 by\time
   \xdef\hourmin{\hourmin:\ifnum\count255<10 0\fi\the\count255}}

\let\a=\alpha \let\b=\beta    \let\g=\gamma     \let\d=\delta     \let\e=\varepsilon
\let\z=\zeta  \let\h=\eta     \let\th=\vartheta \let\k=\kappa     \let\l=\lambda
\let\m=\mu    \let\n=\nu                      \let\r=\rho
\let\s=\sigma \let\t=\tau            \let\c=\chi
\let\ps=\psi   \let\o=\omega     
\let\G=\Gamma        \let\L=\Lambda

\def\PP{{\cal P}}\def\EE{{\cal E}}\def\VV{{\cal V}}
\def\HH{{\cal H}}\def\WW{{\cal W}}
\def\TT{{\cal T}}\def\NN{{\cal N}}
\def\RR{{\cal R}}\def\LL{{\cal L}}
\def\DD{{\cal D}}

\def\xx{{\bf x}}
  
\def\yy{{\bf y}}\def\kk{{\bf k}}\def\nn{{\bf n}}
\def\zz{{\bf z}}
 \def\bP{{\bf P}}

       \def\oo{{\underline \omega}}
\def\ee{{\underline \varepsilon}}

          \def\BBB{\hbox{\euftw B}}
\def\RRR{\hbox{\msytw R}}

\def\NNN{\hbox{\msytw N}}          
        \def\ZZZ{\hbox{\msytw Z}}

        \def\EE{\hbox{\msytw E}}

\let\==\equiv

\let\io=\infty
\let\0=\noindent

\def\*{{\hfill\break\null\hfill\break}}

\def\tilde#1{{\widetilde #1}}

\def\la{{\langle}}
\def\ra{{\rangle}}

\def\tende#1{\,\vtop{\ialign{##\crcr\rightarrowfill\crcr
             \noalign{\kern-1pt\nointerlineskip}
             \hskip3.pt${\scriptstyle #1}$\hskip3.pt\crcr}}\,}
\def\otto{\,{\kern-1.truept\leftarrow\kern-5.truept\to\kern-1.truept}\,}

\def\wh#1{\widehat{#1}}
\def\hat#1{\wh{#1}}
\def\sqt[#1]#2{\root #1\of {#2}}

\def\bp{{\bar \ps}}

\def\PP{{\cal P}}\def\EE{{\cal E}}\def\VV{{\cal V}}
\def\HH{{\cal H}}\def\WW{{\cal W}}
\def\TT{{\cal T}}\def\NN{{\cal N}}
\def\RR{{\cal R}}\def\LL{{\cal L}}
\def\DD{{\cal D}}

\def\T#1{{#1_{\kern-3pt\lower7pt\hbox{$\widetilde{}$}}\kern3pt}}
\def\VVV#1{{\underline #1}_{\kern-3pt
\lower7pt\hbox{$\widetilde{}$}}\kern3pt\,}
\def\W#1{#1_{\kern-3pt\lower7.5pt\hbox{$\widetilde{}$}}\kern2pt\,}

\def\indica{\leaders \hbox to 0.5cm{\hss.\hss}\hfill}
\def\guida{\leaders\hbox to 1em{\hss.\hss}\hfill}
\mathchardef\oo= "0521

\def\xx{{\bf x}}
\def\yy{{\bf y}}\def\kk{{\bf k}}\def\nn{{\bf n}}
\def\zz{{\bf z}}
 \def\bP{{\bf P}}
 
\def\oo{{\underline \omega}}

\def\qed{\raise1pt\hbox{\vrule height5pt width5pt depth0pt}}
 
  \def\bp{{\bar p}} 

\def\indic{\hbox{\raise-2pt \hbox{\indbf 1}}}

\def\RRR{\hbox{\msytw R}} 
 
\def\NNN{\hbox{\msytw N}} 
 \def\ZZZ{\hbox{\msytw Z}}

\def\ins#1#2#3{\vbox to0pt{\kern-#2 \hbox{\kern#1 #3}\vss}\nointerlineskip}

\newdimen\xshift \newdimen\xwidth \newdimen\yshift

\def\insertplot#1#2#3#4#5#6{%
\xwidth=#1pt \xshift=\hsize \advance\xshift by-\xwidth \divide\xshift by 2%
\begin{figure}[ht]
\vspace{#2pt} \hspace{\xshift}
\begin{minipage}{#1pt}
#3 \ifnum\driver=1 \griglia=#6
\ifnum\griglia=1 \openout13=griglia.ps \write13{gsave .2
setlinewidth} \write13{0 10 #1 {dup 0 moveto #2 lineto } for}
\write13{0 10 #2 {dup 0 exch moveto #1 exch lineto } for}
\write13{stroke} \write13{.5 setlinewidth} \write13{0 50 #1 {dup 0
moveto #2 lineto } for} \write13{0 50 #2 {dup 0 exch moveto #1
exch lineto } for} \write13{stroke grestore} \closeout13
\includegraphics{griglia.ps} \fi
\includegraphics{#4.ps}\fi%
\ifnum\driver=2 \fi
\end{minipage}
\caption{#5}
\end{figure}
}

\newdimen\shift \shift=-1.5truecm
\def\lb#1{%
\ifnum\bozza=1
\label{#1}\rlap{\hbox{\hskip\shift$\scriptstyle#1$}}
\else\label{#1} \fi}

\def\be{\begin{equation}}
\def\ee{\end{equation}}
\def\bea{\begin{eqnarray}}\def\eea{\end{eqnarray}}
\def\bean{\begin{eqnarray*}}\def\eean{\end{eqnarray*}}
\def\bfr{\begin{flushright}}\def\efr{\end{flushright}}
\def\bc{\begin{center}}\def\ec{\end{center}}
\def\bal{\begin{align}}\def\eal{\end{align}}
\def\ba#1{\begin{array}{#1}} \def\ea{\end{array}}
\def\bd{\begin{description}}\def\ed{\end{description}}
\def\bv{\begin{verbatim}}\def\ev{\end{verbatim}}
\def\nn{\nonumber}
\def\Halmos{\hfill\vrule height10pt width4pt depth2pt \par\hbox to \hsize{}}
\def\pref#1{(\ref{#1})}
 
\def\virg{\quad,\quad}

\def\ins#1#2#3{\vbox to0pt{\kern-#2 \hbox{\kern#1 #3}\vss}\nointerlineskip}

\newdimen\xshift \newdimen\xwidth \newdimen\yshift
\newcount\griglia

\def\insertplot#1#2#3#4#5#6{%
\xwidth=#1pt \xshift=\hsize \advance\xshift by-\xwidth \divide\xshift by 2%
\begin{figure}[ht]
\vspace{#2pt} \hspace{\xshift}
\begin{minipage}{#1pt}
#3 \ifnum\driver=1 \griglia=#6
\ifnum\griglia=1 \openout13=griglia.ps \write13{gsave .2
setlinewidth} \write13{0 10 #1 {dup 0 moveto #2 lineto } for}
\write13{0 10 #2 {dup 0 exch moveto #1 exch lineto } for}
\write13{stroke} \write13{.5 setlinewidth} \write13{0 50 #1 {dup 0
moveto #2 lineto } for} \write13{0 50 #2 {dup 0 exch moveto #1
exch lineto } for} \write13{stroke grestore} \closeout13
\includegraphics{griglia.ps} \fi
\includegraphics{#4.ps}\fi%
\ifnum\driver=2 \fi
\end{minipage}
\caption{#5}
\end{figure}
}

\newdimen\shift \shift=-1.5truecm
\def\lb#1{%
\label{#1}\rlap{\hbox{\hskip\shift$\scriptstyle#1$}}
\else\label{#1} \fi}

\def\be{\begin{equation}}
\def\ee{\end{equation}}
\def\bea{\begin{eqnarray}}\def\eea{\end{eqnarray}}
\def\bean{\begin{eqnarray*}}\def\eean{\end{eqnarray*}}
\def\bfr{\begin{flushright}}\def\efr{\end{flushright}}
\def\bc{\begin{center}}\def\ec{\end{center}}
\def\bal{\begin{align}}\def\eal{\end{align}}
\def\ba#1{\begin{array}{#1}} \def\ea{\end{array}}
\def\bd{\begin{description}}\def\ed{\end{description}}
\def\bv{\begin{verbatim}}\def\ev{\end{verbatim}}
\def\nn{\nonumber}
\def\Halmos{\hfill\vrule height10pt width4pt depth2pt \par\hbox to \hsize{}}
\def\pref#1{(\ref{#1})}
 
\def\virg{\quad,\quad}

\usepackage{amsmath}
\usepackage{amsfonts}
\usepackage{amssymb}
\usepackage{epstopdf}

\font\msytw=msbm9 scaled\magstep1 
scaled\magstep1

\let\a=\alpha \let\b=\beta  \let\g=\gamma  \let\d=\delta
\let\e=\varepsilon
\let\z=\zeta  \let\h=\eta   \let\th=\theta \let\k=\kappa \let\l=\lambda
\let\m=\mu    \let\n=\nu             \let\r=\rho
\let\s=\sigma \let\t=\tau    \let\c=\chi
\let\ps=\Psi   \let\o=\omega
\let\G=\Gamma   \let\L=\Lambda

\def\EE{{\cal E}} \def\VV{{\cal V}}
 \def\WW{{\cal W}}
\def\TT{{\cal T}}\def\NN{{\cal N}} \def\BBB{{\cal B}}
\def\RR{{\cal R}}\def\LL{{\cal L}}  
\def\DD{{\cal D}}

 \def\xx{{\bf x}} \def\yy{{\bf y}} \def\zz{{\bf z}}
\def\kk{{\bf k}}
\def\PP{{\bf P}}

\def\nn{\nonumber}

\def\RRR{\hbox{\msytw R}} 

\def\NNN{\hbox{\msytw N}} 
 \def\ZZZ{\hbox{\msytw Z}}

\def\\{\hfill\break}
\def\={:=}
\let\io=\infty
\let\0=\noindent

\def\tende#1{\,\vtop{\ialign{##\crcr\rightarrowfill\crcr\noalign{\kern-1pt
    \nointerlineskip} \hskip3.pt${\scriptstyle #1}$\hskip3.pt\crcr}}\,}
\def\otto{\,{\kern-1.truept\leftarrow\kern-5.truept\to\kern-1.truept}\,}

\def\wh{\widehat}
\def\to{\rightarrow}
\def\la{\left\langle}
\def\ra{\right\rangle}
\def\qed{\hfill\raise1pt\hbox{\vrule height5pt width5pt depth0pt}}

\def\be{\begin{equation}}
\def\ee{\end{equation}}
\def\bp{\begin{pmatrix}}
\def\ep{\end{pmatrix}}
\def\bea{\begin{eqnarray}}
\def\eea{\end{eqnarray}}
\def\nn{\nonumber}
\def\pref#1{(\ref{#1})}

\def\lb{\label}

\newtheorem{lemma}{Lemma}[section]
\newtheorem{theorem}{Theorem}[section]

\begin{document}

\title{Localization in an interacting quasi-periodic fermionic
chain
}

\author{Vieri Mastropietro}

\affiliation{
Universit\'a di Milano, Via C. Saldini 50, 20133, Milano, Italy
}

\begin{abstract}
We consider a many body fermionic system with an incommensurate
external potential and a short range interaction in one dimension. 
We prove that, for certain densities and weak interactions, 
the zero temperature thermodynamical correlations are exponentially
decaying for large distances, a property indicating persistence of
localization in the interacting ground state. The analysis is based on
Renormalization Group, and convergence of the renormalized expansion
is achieved using fermionic
cancellations and controlling the small divisor problem assuming a
Diophantine condition for the frequency. 
\end{abstract}


\maketitle

\renewcommand{\thesection}{\arabic{section}}

\section{Introduction and main results}

\subsection{Introduction}

The properties of a fermionic system, like the conduction
electrons in a metal, are determined, when the interaction between
particles is not taken into account, by the eigenfunctions of the
single particle hamiltonian. In presence of an external periodic
potential, the eigenfunctions are Bloch waves, and the zero
temperature a.c. conductivity is vanishing (insulating behavior) or
not (metallic behavior) depending if the Fermi level lies in
correspondence of a gap in the single particle spectrum or not. A
different way in which an external potential can produce an
insulating behavior is known as {\it Anderson localization} \cite{A}; in presence of certain potentials
(like random ones, physically describing the presence of unavoidable impurities in the metal)
the eigenfunctions of the single particle Hamiltonian can be
exponentially localized and this produces an insulating behavior. 
Localization in the single particle Schroedinger equation with a random field has been indeed 
rigorously
proved in various regimes of energy and disorder, starting from \cite{loc0},\cite{loc1}. Note that in one
dimension typically any amount of disorder produces localization,
while in three dimensions the disorder has to be sufficiently
strong and a metal to insulator transition is expected varying the
strength of the random field. Localization does not necessarily require disorder, as it has
long been known \cite{AA}
that also nonrandom systems with quasi-periodic potentials (or incommensurate in the lattice case) can present single particle localization.
The one dimensional quasi-periodic Schroedinger equation has {\it extended} Bloch-Floquet eigenfunctions
in the weak coupling regime
\cite{DS},\cite{E} and {\it localized} eigenfunctions 
in the strong coupling regime, see
\cite{BLT},\cite{S},\cite{FS},\cite{Si}, provided that some Diophantine condition is assumed.
In the lattice case with a
cosine potential, the weak or strong
coupling regime are connected by a duality transformation
\cite{AA}, and in this case it can be proved \cite{Av} that the
spectrum is a Cantor set. In any case, the case of 1D quasi-periodic potential resembles the 3D random case, 
as there is a transition between an extended and localized phase varying the strength of the potential.

A realistic description of metals must include
the electron-electron interaction, so that the problem of the interplay between localization and interactions naturally arises
\cite{FA}. 
In the physical literature
the zero temperature thermodynamical properties of 1D interacting fermions with disorder has been analyzed in 
\cite{GT}, \cite{ GGG}, finding 
localized and delocalized regions; the quasi-periodic case
have been studied in
\cite{VMG}.
While such works 
concern the computation of the zero temperature thermodynamical quantities, in more recent times attention has been devoted also to
the localization properties of excited states of interacting disordered many body systems, see
 \cite{B},\cite{PH}. 
Evidence has been found that in several interacting systems with disorder
all the eigenfunctions are localized for weak interactions, while stronger interactions can destroy localization, leading to a so-called many-body localization transition; similar properties 
has been found also in the quasi-periodic case \cite{NH}, \cite{H4}.

It should however remarked that not only the results about 
the excited states of the N-particle Hamiltonian but even the ground state properties (that is, the
zero temperature thermodynamical quantities)
are based on 
conjectures or approximations, and more quantitative results based on rigorous methods seem necessary.
In particular, there is still no a rigorously established example of an interacting many body system in which localization, which was present in the non interacting single particle state, still persists in presence of interaction.
The mathematical tools used for single particle localization in the disordered case can actually treat the case
of a {\it finite} number of interacting disordered particles, see \cite{loc2}.
By using KAM or block Jacobi procedure, localization of most eigenstates (in the sense that the expectations of local observables are exponentially decaying)
has been rigorously proved in 
\cite{loc4} (see also \cite{loc4a}) in a many body interacting disordered fermionic chain , 
under a physically reasonable {\it assumption}
that limits the amount of level attraction in the system.   
Evidence of localization for finite times in interacting disordered bosons has been found in 
\cite{loc3}. 

There exist powerful methods, based on the version of Renormalization Group (RG)
developed for constructive Quantum Field Theory, to compute the thermodynamical properties  
at zero temperature of interacting fermions. Such techniques encounter at the moment some difficulty in
the application to random fermions, but can be successfully applied in the case quasi-periodic or incommensurate
potentials; this is not surprising as  quasi-periodic potentials produce {\it small divisors} 
similar to the ones in the KAM Lindstedt series, whose convergence can be established by RG methods, see \cite{G},\cite{GM1}. We will therefore analyze the
interplay of localization and interaction in the thermodynamical functions 
of interacting fermions with a quasi-periodic potential by RG methods.
In particular, if $\L$ is a one dimensional
lattice $\L=\{x\in \mathbb{Z}
, -[L/2]\le x\le [(L-1)/2]\}$, we consider a system of fermions with Hamiltonian
\be H_N=-\e\sum_{i=1}^N \tilde\partial^2_{x_i}+u\sum_{i=1}^N \bar
\phi_x+\l \sum_{i,j=1\atop i\not=j}^N v(x_i-x_j)\label{ass1} \ee
where $\tilde\partial^2_{x} f(x)=f(x+1)+f(x-1)-2 f(x)$,
$\phi_x=\bar\phi(
\o x)$ with $\bar\phi(t)=\bar\phi(t+1
)$, $\o$ irrational
and $v(x-y)=\d_{y,x+1}$ for definiteness. When $\phi_x=\cos (\o x 2\pi)$ 
and $\l=0$ the above model is the
interacting version of the Aubry-Andr\'e model \cite{AA}. 
In absence of interactions between particles $\l=0$ the
eigenfunctions of $H_N$ are the antisymmetric product of the
single particle eigenfunctions of
Schroedinger equation
\be -\e\psi(x+1)-\e\psi(x-1)+u \phi_x \psi(x)=E\psi(x)\label{se}
\ee
which were extensively analyzed \cite{DS},\cite{E}, \cite{BLT},\cite{S},\cite{FS},\cite{Si}.
In principle, the thermodynamical quantities could be obtained from such studies but, as a matter of fact, 
even in the $\l=0$ case the only available results on the zero temperature properties of \pref{ass1} were obtained by RG methods for functional integrals.
Indeed in \cite{BGM} 
the Grand canonical correlations with $\l=0$ were written in terms of an expansion
plagued by a small divisor
problem, and convergence was proved in \cite{BGM}, for small  ${u\over \e}$, suitable chemical potentials and assuming a Diophantine condition on the frequencies, that is
$
||2\pi n\o||_{2\pi}\ge C
n^{-\t}\quad {\rm for} \quad {\rm any} \quad n\in  \mathbb{Z}/\{0\}$
 where $||.||_{2\pi}$ is the norm on the one dimensional
torus with period $2\pi$. It was found a power law or an exponential decay of the zero temperature correlations at large distances
depending if the chemical potential is inside a gap or not; that is metallic or a band insulator behavior. In the opposite limit
when $u/\e$ is {\it large} in the non interacting case $\l=0$
it was proved in \cite{GM} that the correlations decay exponentially for suitable values
of the chemical potential, in agreement with the localization properties
of the single particle eigenfunction. 

The only rigorous result for quasi-periodic {\it interacting} fermions
is in \cite{M}, in which it was proved that
for small ${u\over \e}$ and small $\l$ there is still a power law decay of correlations
for values of the chemical potential outside the gap, but the exponent is anomalous
with a critical exponent signaling Luttinger liquid behavior. Therefore the metallic behavior, which was present in the non interacting case as consequence of the extended nature of the single particle eigenfunctions, persists also in presence of interaction (but one has a Luttinger liquid instead than a Fermi liquid).
In addition if the chemical potential is inside a gap one has exponentially decay of correlation, and
an anomalous exponent appears 
in the decay rate.  

In the  present paper we finally consider a system of {\it interacting} fermions
with a {\it large} incommensurate potential, and we prove 
that, for suitable chemical potentials, 
the zero temperature thermodynamical correlations are exponentially
decaying for large distances for weak interaction,
a property indicating persistence of
localization in the interacting ground state.

\subsection{Thermodynamical quantities and solvable limits}


We consider the Grand-canonical ensemble, in which one performs averages over the particle number.
One introduces fermionic creation and annihilation operators  $a^+_{x},a^-_{x}$ on the Fock space verifying 
$\{a^{\e}_x,a^{-\e'}_y\}=\d_{\e,\e'}\d_{x,y}$. The Fock space Hamiltonian corresponding to 
\pref{ass1}
can be written as  
\be H=-\e\sum_{x\in\L} (a^+_{x+ 1} a_{x}+a^+_{x} a^-_{x+1} )+
u\sum_{x\in\L} \bar \phi(\o x)  a^+_{x} a^-_{x}-\m\sum_{x\in\L}
a^+_{x} a^-_{x}+\l \sum_{x\in\L}a^+_{x} a^-_{x} a^+_{x+1}
a^-_{x+1}\label{1.1} \ee
%
Using the Jordan-Wigner transformation the
model can be mapped in the XXZ model with a coordinate dependent
magnetic field $h_x=\phi_x$. 
%
%


Let us consider now the thermodynamical quantities in the grand-canonical ensemble.
We consider the operators $a_{\xx}^{\pm}=e^{x_0 H}
a_x^{\pm}e^{-H x_0}$, with
\be \xx=(x,x_0)\;,\quad 0\le x_0 < \b\ \ee
for some $\b>0$ ($\b^{-1}$ is the temperature); $x_0$ is the
imaginary time and on it antiperiodic boundary conditions are
imposed, that is, if $a_{\xx}^{\pm}=a_{x,x_0,s}^{\pm}$, then
$a_{x,\b}^{\pm}=-a_{x,0}^{\pm}$.
The 2-point {\it Schwinger function} is defined as
\be\lb{1.6} {{\rm Tr} \left[e^{-\b H} {\bf T} (
a^-_\xx a^+_\yy)\right] \over {\rm Tr} [e^{-\b H_0}]}=
{\bf I}(x_0-y_0>0){{\rm Tr}[e^{-\b H} a^-_\xx a^+_\yy
]\over
{\rm Tr}[e^{-\b H}]}-{\bf I}(x_0-y_0\le 0) {{\rm Tr}[e^{-\b H} a^+_\yy a^-_{\yy}
]\over
{\rm Tr}[e^{-\b H}]}
\ee
where ${\bf T}$ is the time order product.
The above quantity cannot be exactly computed, so that one has
to rely on a perturbative expansion around some solvable limit. In particular the model is solvable in the {\it free fermion} limit ($\l=u=0)$, which is an extended phase
and in the {\it molecular limit}, which is a localized phase; in order to investigate
the interplay of localization and interaction we will perform an expansion around the molecular limit.
Before doing that, let us discuss the main properties
of the solvable limits.

In the {\it free fermion limit}, corresponding to $u=\l=0$,
the Hamiltonian can be written in diagonal form in
momentum space. If we assume periodic boundary conditions and we
set $a_x^\pm={1\over L}\sum_{k}e^{\pm i k x} \hat a^\pm_k$, with
$k={2\pi \over L}n$ and $\{\hat a^{\e},\hat
a^{-\e'}_{k'}\}=\d_{\e,\e'}\d_{k,k'}$ then ($\e=1$ for definiteness)
\be
H_0=\sum_k (-\cos k+\m)\hat a^{+}_{k} \hat a^-_{k}
\ee
%
The two points $\pm p_F$ are called {\it Fermi points}.
The two point Schwinger function is equal to
\bea\label{defo}
&&G(\xx-\yy)= {{\rm Tr} \left[e^{-\b H_0} {\bf T} (
a^-_\xx a^+_\yy)\right] \over {\rm Tr} [e^{-\b H_0}]} =
{1\over L} \sum_{k} e^{-ik(x-y)} \hat G(k,x_0-y_0)=\\
&&={1\over L} \sum_{k} e^{-ik(x-y)} \left\{ {e^{-(x_0-y_0) e(k)}
\over 1+e^{-\b e(k)}}{\bf I}(x_0-y_0>0) - {e^{-(\b+x_0-y_0) e(k)} \over
1+e^{-\b e(k)}}{\bf I}(x_0-y_0\le 0) \right\}
\nn\eea
where $\e(k)=\m-\cos k$ 
The function $\hat G(k,\t)$ is defined only for $-\b<\t< \b$, but we can
extend it periodically over the whole real axis. 
The function $\hat G(k,\t)$ is antiperiodic in $\t$ of period $\b$; hence
its Fourier series is of the form
\be
\hat G(k,\t)={1\over\b}\sum_{k_0={2\pi\over\b}(n_0+{1\over 2})}
\hat G(k_0,k)e^{-i k_0 \t}\ee
with
\be \hat G(k,k_0)=\int_0^\b d\t e^{i \t k_0}{e^{-\t e(k)} \over 1+e^{-\b
e(k)}}= {1\over -i k_0+e(k)}\label{l1} \ee
Note that the function $\hat G(\kk)$ is singular, in the limit
$L\to\io,\b\to\io$, at $k_0=0, k=\pm p_F$, with  
$\cos p_F=\m$. $\pm p_F$ are the Fermi moments 
and close to them, that is for $k'$ small it behaves as
\be \hat G(k'\pm p_F,k_0)\sim {1\over -i k_0\pm v_F k'}\label{l2}
\ee
%

Another solvable limit is the
{\it Molecular limit} corresponding to $\l=\e=0$.
The Hamiltonian reduces to ($u=1$ for definiteness)
\be\lb{1.1aa}  H_0=\sum_{x\in\L}(\phi_x-\m)a^+_{x}a^-_{x}\ee
%
%
%
%
The $2$--point function $g(\xx,\yy)=\la{\bf
T}\{a^-_{\xx} a^+_{\yy}\}\ra_{\b,L}$ is equal to
\be\label{defo}
g(\xx,\yy)=\d_{x,y}\left\{ {e^{-(x_0-y_0) (\phi_x-\m)}
\over 1+e^{-\b (\phi_x-\m)}}{\bf I}(x_0-y_0>0) - {e^{-(\b+x_0-y_0)(\phi_x-\m)} \over
1+e^{-\b (\phi_x-\m)}}{\bf I}(x_0-y_0\le 0)\right\} =\d_{x,y}\bar g (x,x_0-y_0)\ee
The function $\bar g(x,\t)$ is defined only for $-\b<\t< \b$, but we can
extend it periodically over the whole real axis. This periodic
extension is smooth in $\t$ for
$\t\not= n\b, n\in \mathbb{Z}$, but has a jump discontinuity at $\t=n\b$
equal to $(-1)^n$.

The function $g(\xx,\yy)$ is antiperiodic in $x_0-y_0$ of period $\b$; hence
its Fourier series is of the form
\be
g(\xx,\yy)=\d_{x,y}{1\over\b}\sum_{k_0={2\pi\over\b}(n_0+{1\over 2})}
\hat g(x,k_0)e^{-i k_0 (x_0-y_0)}\ee
with
\be \hat g(x,k_0)=\int_0^\b d\t e^{i \t k_0}{e^{-\t (\phi_x-\m)} \over 1+e^{-\b
 (\phi_x-\m)
}}= {1\over -i k_0+\phi_x-\m} \ee
Let $M\in\NNN$ and $\c(t)$ a smooth compact
support function that is $1$ for $t\le 1$ and $0$ for $t\ge \g$, with 
$\g>1$. 
Let $\DD_{\b}=D_{\b}\cap\{k_0\,:\,\c_0(\g^{-M}|k_0|)>0\}$,
where $D_\b=\{k_0={2\pi\over\b}(n_0+{1\over 2}), n_0\in \ZZZ\}$.
If $x_0-y_0\not= n\b$, we can write
\be\lb{limN} g(\xx,\yy) = \lim_{M\to \io} \d_{x,y}\frac1{\b} \sum_{k_0\in \DD_\b}\chi(\g^{-M}|k_0|)
\frac{e^{-i k_0(x_0-y_0)}}{-i k_0+\phi_x-\m}\equiv \d_{x,y}{1\over\b}
\sum_{k_0\in
D_{\b}}e^{-i k_0(x_0-y_0)}\hat g^{(\le M)}(x,k_0)\equiv  \lim_{M\to \io} g^{(\le M)}(\xx,\yy) \ee
Because of the jump discontinuities, $g^{(\le M)}(\xx,\yy)$ is not absolutely
convergent but 
is pointwise convergent
and the limit is given by $g(\xx,\yy)$
at the continuity points, while at the discontinuities it is given by
the mean of the right and left limits. 
%
%

In particular, the above equality is not true for $x_0-y_0=n\b$, where
the propagator is equal
\be
\bar g(x,0^-)\to_{\b,L\to\io} -{\bf I}(\phi_x-\m\le 0)
\ee
while
the r.h.s. is equal to
\be\lb{1.26}{ \bar g(x,0^-)+ \bar g(x,0^+)\over 2}\to_{\b,L\to\io}[
{1\over 2}-{\bf I}(\phi_x-\m\le 0)]
\ee
Assuming that $\bar\phi(x)$ is even, periodic $\bar\phi(x)=\bar\phi(x+1)$ and that
there is only one $\bar x
\in (0, {1\over 2})$ such that
$\m=\bar\phi(\bar \o)$; setting $x=x'\pm \bar x$ we have that, for small $\o x'$ (mod. 1)
\be
\phi_{x'+\r  \bar x}-\m=\r v_0 \o x'+r_{x'}\quad v_0=\partial \bar\phi(\o\bar x),\quad \r=\pm
\ee
therefore the 2-point function can be written as
\be
\hat g(x'\pm \bar x, k_0)\sim {1\over -i k_0\pm v \o x'}
\label{l3}
\ee
Note the similarity of \pref{l3} with \pref{l2}; this analogy suggests to call $\pm \bar x$ as {\it Fermi coordinates}, in analogy with the Fermi momenta $\pm p_F$.
In the special case of $\phi_x=\cos(2\pi\o x)$ (Almost-Mathieu operator),  setting $\e=u$
\be
\hat G(k,k_0)|_{k=2\pi \o x}=\hat g(x,k_0)
\ee
which is is a manifestation of the well known {\it Aubry-duality}.

\subsection{Grassmann Integral representation}

If
$\BBB_{\b,L}=\{\DD_\b\bigcup\L\}$, we consider the
Grassmann algebra generated by the Grassmannian variables
$\{\psi^\pm_{x,k_0}\}_{ x, k_0 \in
\BBB_{\b,L}}$ and a Grassmann
integration $\int
\big[\prod_{x,k_0\in\BBB_{\b,L}}
d\psi_{x,k_0}^- d\psi_{x,k_0}^+\big]$ defined as
the linear operator on the Grassmann algebra such that, given a
monomial $Q(\psi^-, \psi^+)$ in the variables
$\psi^\pm_{x,k_0}$, its action on $Q(\psi^-,
\psi^+)$ is $0$ except in the case $Q(\psi^-,\psi^+)=\prod_{x,k_0\in\BBB_{\b,L}}
\psi^-_{x,k_0}
\psi^+_{x,k_0}$, up to a permutation of the variables. In
this case the value of the integral is determined, by using the
anticommuting properties of the variables, by the condition
\be \int
\Big[\prod_{x,k_0\in\BBB_{\b,L}}
d\psi_{x,k_0}^+
d\psi_{x,k_0}^-\Big]\prod_{x,k_0\in\BBB_{\b,L}}
\psi^-_{x,k_0}
\psi^+_{x,k_0}=1\label{2.1}\ee
We define also Grassmanian field as 
$\psi^\pm_\xx={1\over \b}\sum_{k_0\in \DD_\b} e^{\pm i k_0  x_0}\psi^\pm_{x,k_0}$ 
with $x_0=m_0 {\b\over \g^M}$ and $m_0\in (0,1,...,\g^M-1)$.
The "Gaussiam Grassmann measure"  is defined as
\be
P(d\psi)=[\prod_{x,k_0\in\BBB_{\b,L}}\b
d\psi_{x,k_0}^- d\psi_{x,k_0}^+  
\hat g^{(\le M)}(x,k_0)]\exp\{ -\sum_{x,k_0} (\hat g^{(\le M)}(x,k_0))^{-1}\psi^+_{x,k_0}\psi^-_{x,k_0}   \} \label{ch}
\ee
%
We introduce the generating functional $W_M(\phi)$ defined in terms of the following Grassmann integral (free boundary conditions in space are assumed)
\be
e^{W_M(\phi)}=\int P(d\psi)e^{-\VV^{(M)}(\psi)- \BBB^{(M)}(\psi,\phi) )
}\label{GI}
\ee
where $\psi^\pm_{\xx}$ and $\phi^\pm_{\xx,s}$ are Grassmann variables, $P(d\psi)$ has propagator
\be
g^{(\le M)}(\xx,\yy)=
\d_{x,y}\frac1{\b} \sum_{k_0\in \DD_\b}\chi(\g^{-M}|k_0|)
\frac{e^{-i k_0(x_0-y_0)}}{-i k_0+\phi_x-\phi_{\bar x}}\label{pro11}
\ee
and
$\int d\xx$ is a short form for $\sum_{x\in\L}{\b\over\g^M}\sum_{x_0}
$;
%
%
moreover
\bea
&&\VV^{(M)}=\l\int d\xx \psi^{+}_{\xx}\psi^{-}_{\xx} \psi^{+}_{\xx+{\bf e_1}}\psi^{-}_{\xx+{\bf e_1}}
+\e \int d\xx [\psi^{+}_{\xx+{\bf e_1}}\psi^{-}_{\xx}+ \psi^{+}_{\xx}\psi^{-}_{\xx+{\bf e_1}}]\nn\\
&& +\n\int d\xx\label{VM}
 \psi^{+}_{\xx}\psi^{-}_{\xx}+\int d\xx \n_C(x+1)
 \psi^{+}_{\xx}\psi^{-}_{\xx}+\int d\xx \n_C(x)
 \psi^{+}_{\xx+{\bf e_1}}\psi^{-}_{\xx+{\bf e_1}}\nn
\eea
where
\be\lb{lll} \n_C(x)={1\over 2} \l
[\bar g(x,0^+)-\bar g(x,0^-)]
\ee
and $\bar g(x,0^-)$ was defined in \pref{defo}.
Finally
\be
\BBB^{(M)}(\psi,\phi) =   
\int d\xx [\phi^+_{\xx} \psi^-_{\xx} + \psi^+_{\xx}
\phi^-_{\xx}]\ee
Note that we expect that the chemical potential is modified by the interaction; in the analysis it is convenient to keep fixed the value of the Fermi coordinate in the free or interacting theory,
therefore we write the chemical potential as $\phi_{\bar x}+\n$, where $\n$ is a counterterm to be fixed so that the free and interacting Fermi coordinate are the same.

Let us define
\be
S_{2}^{M,\b,L}(\xx,\yy)
={\partial^2\over\partial\phi^+_{\xx}\partial\phi^{-}_{\yy}}
W_M(\phi)|_{0}\label{asso}
\ee
Note that $\lim_{M\to \io}S_{2}^{M,\b,L}$ can be written as a series in $\e,\l$
coinciding order by order with the series expansion for 
the Schwinger functions \pref{1.6} with chemical potential $\m=\phi_{\bar x}+\n$.
Indeed each term
of the series for \pref{1.6} or $\lim_{M\to \io}S_{2}^{M,\b,L}$
can be expressed as a sum of integrals over propagators (respectively
$g(\xx,\yy)$ \pref{defo}
or $\lim_{M\to\io} g^{(\le M)}(\xx,\yy)$ \pref{ch})
which can be represented by
Feynman graphs.
The subset of graphs contributing to \pref{1.6}
and with no tadpoles coincides the the graphs contributing to
$\lim_{M\to\io} S_{2}^{M,\b,L}$ and no vertices $\n_C$.
The integrands are different, as the propagators
$g(\xx,\yy)$ \pref{defo}
or $\lim_{M\to\io} g^{(\le M)}(\xx,\yy)$ \pref{ch})
are different at coinciding times. However the integrals are well defined and coincide, as
the integrands of the graphs coincide
except in a
set of zero measure. Let us consider the remaining graphs.
In the graphs with a tadpole in the
expansion for
$\lim_{M\to\io} S^{M,\b,L}_2$ there is
a factor of the form
\be
g(\xx_1-\xx) \n_T(x+1)) g(\xx-\xx_2) \virg
\n_T(x+1) = -{\l\over 2 }[\bar g(x+1,0^+)+\bar g(x+1,0^-)]
\ee
On the other hand, given a graph
$G$ of this type, there is another graph $\tilde G$, which differs from it
only because, in place of the term $\VV(\psi)$ which produced the tadpole,
there is a vertex $\n_C(x+1)$. If we sum the values of $G$ and $\tilde
G$, we get a number which is equal to the value of $G$, with $
 -\l\bar g(x+1,0^-)$ replacing $\n_T(x+1)$
, so that the terms coincide with the analogous term in the expansion for
\pref{1.6}. Therefore the perturbative expansion coincide. 
An analyticity argument, analogue to the one in  
Proposition  2.1 of \cite{BFM}, would allow to conclude the coincidence of 
\pref{1.6} and $\lim_{M\to \io}S_{2}^{M,\b,L}$ beyond perturbation theory, once that the limit exists
and certain analyticity properties are proved; this is quite standard and 
will be not repeated here for brevity, so we state our main results directly for the Grassmann integral.

\subsection{Main result: localization in presence of interaction}

We set $u=1$ and we consider $\e,\l$
small. We define $\mathbb{T}=\mathbb{R}/\mathbb{Z}$ the one dimensional torus,
$||\th||_{1}$  the norm, that is the absolute value of
$\th$ modulo 1 defined so that $0\le |\th|_1\le {1\over 2}$. 
Our main result is the following.

\begin{theorem}
Let us consider $\phi_x=\bar
\phi(\o x)$ an even function in $C^1(\mathbb{T})$, that is $\bar\phi(x)=\bar\phi(x+1)$
and $\phi_x=\phi_{-x}$, with $|\phi_x|\le 1$ and $\phi(t)$ increasing for $0<t<{1\over 2}$.
We consider the 2-point function \pref{asso}
with $\bar x$ half integer and so that 
$\partial_x \bar\phi(\bar x)>0$. Assume \be ||\o
x||_1\ge C |x|^{-\t}, \quad {\rm for} \quad {\rm any} \quad 0\not
=x\in  \mathbb{Z}\label{dc} \ee 
%
For $\e$ small and $|\l|
\le \e^{2\bar x+2}$ there exists a continuous function $\n(\e,\l)$ such that, for any $N$, the limit
$\lim_{\b\to\io}\lim_{L\to\io}\lim_{M\to\io} S^{M,\b,L}_2(\xx,\yy)=S_2(\xx,\yy)$ exists and verifies, for any $N\in  \mathbb{N} $
\be
|S_2(\xx,\yy)|\le C_N {e^{-\k \log|\e|^{-1}|x-y|}\over 1+(|\s||x_0-y_0|)^N}
\label{xx1}
\ee
where $\s=O(\e^{2\bar x})$ and non vanishing and  $\k,C_N$ positive constants .
%
%
%
\end{theorem}

The above theorem is proved by an expansion in $\l,\e$ around the molecular limit, considering the kinetic energy and and the many body interaction as perturbations, and assuming the Fermi coordinate $\bar x$
as equal to a half integer.
If there is no interaction $\l=0$, 
the exponential decay of the two point function is in agreement with the localization 
of the single particle eigenfunctions of the Schroedinger equation, see for instance lemma 4.3 of
\cite{FS}.  The above theorem says that the
exponential decay persists in presence of interaction, for certain chemical potentials
provided that the hopping is
smaller than $O(\bar x!^{-\g})$ for some positive $\g$, and the interaction is
much smaller than the hopping. As the Grand-canonical averages
reduces to the average over the ground state, such result
indicates localization for the ground state eigenfunction of
an interacting many body system.

A consequence of Theorem 1.1 combined with \cite{M}
is the existence of a {\it quantum phase transition}
between an extended and a localized phase. Indeed it was proved in \cite{M}
in the small $\l,u$ case that even in presence of interaction the system has a metallic or a band insulating behavior; that is, for small $u$ and $\l$ ($\e=1$)
if
$\m=1-\cos p_{F}$ then 
if 
$p_{F}=m\o\pi$ then $S_2(\xx,\yy)$ decays faster than any power (band insulator behavior) with rate
$|\s|=\hat\phi_m |u|^{1-\h}$
where $\h=a\l+O(\l^2)$ with $a>0$ suitable constant; while if 
$|| 2p_{F}+2\pi
n \o||_{2\pi}\ge C |n|^{-\t}, \quad {\rm for} \quad {\rm
any} \quad x\in  \mathbb{Z}/\{0\}$
 then it decays a a power law as $O(|\xx-\yy|^{-1-\h})$ (metallic behavior)
with $\h=b\l^2+O(\l^3)$, $b$ a positive constant. Therefore, in presence of interaction increasing the amplitude of the quasi-periodic potential one moves from an extended to a localized phase.

\subsection{Sketch of the proof of Theorem 1.1 and contents}

In order to prove Theorem 1.1 one has to face a small divisor problem
resembling the one in 
KAM Linstedt series \cite{G}; its origin
lies in the fact that the expansion is in terms of sum of product of propagators $(-i k_0+\phi_x-\phi_{\bar x})^{-1}$, and, 
due to the irrationality of $\o$, propagators with very different $x$ can be very close.  
There are however essential differences with respect to KAM Lindstedt series
or in the non interacting $\l=0$ case;
in such cases
the series can be represented in terms of 
{\it tree} diagrams, while in the present case the series are expressed in terms of {\it diagrams with loops}.
The number of tree diagrams contributing to order $n$ in the perturbative expansion is $O(n!)$ and a ${C^n\over n!
}$-bound on each diagram is sufficient for convergence 
; in presence of interaction, on the contrary,  
the number of diagrams $O(n!^2)$ and a similar bound on each diagram is not sufficient to achieve convergence. 
One has therefore to combine methods developed in the context of KAM with constructive Quantum Field theory techniques; in particular one  has to use the fact that the fermionic expectations can be represented in terms of determinants. 

We perform the analysis of the Grassmann integral \pref{asso} in an iterative way by using Renormalization Group methods. We start integrating the higher energy frequencies, see \S 2.
Here there is not a small divisor problem but one has to show that the expansions are convergent using that the expansion can be written in term of Gram bounds.
After the integration of the ultraviolet fields, we have to integrate the low energy modes (infrared scales) in which
one has to face a small divisor problem, as discussed in \S 3. 
The theory is {\it non-renormalizable} according to power counting; the scaling dimension depends on 
the number of vertices in the subgraph, so that one has to improve the dimensions of all possible subgraphs
with any number of external fields. In order to get such improvement, we have to exploit the incommensurability of the potential and take advantage from the diophantine condition on the frequency.
One has to distinguish between two kind of terms in the effective potential, depending if the coordinates (measured from the Fermi coordinate) of the external fields are different ({\it non-resonant} terms)
or equal ({\it resonant} terms). In the non resonant terms one uses the Diophantine condition
to get good bounds, exploiting, roughly speaking, the idea that if the denominators  associated to the external lines have similar small size but different coordinates, then the difference of coordinates is necessarily large (see lemma 3.2, 3.3 and 3.4 in \S 3.C). The result is somewhat similar to Bruno lemma as presented in \cite{G}, but new difficulties raise from the fact that the resonances have any number of external fields and not only two as in the non interacting case; in particular, one has to improve the bounds by a quantity proportional to the external lines for combinatorial reason, see \S 3.D . 
Regarding the resonances one uses that the local part of the terms with more than four external fields is vanishing. Moreover
the resonances with two external fields 
produce a mass term implying an exponential decay in time; in particular
the propagators associated to the two external fields have coordinate $x=x+n$
and
$\phi_{\bar
x}=\bar\phi_{\bar x+n}$ either when $n=0$ {\it or} $n=-2 \bar x$. The
second case is responsible of the mass term while the first case produces the renormalization of the Fermi coordinates. Finally in \S 3.F we study also the flow of the running coupling constants
and the two point functions, 
completing the theorem proof.

\section{The ultraviolet integration}

\subsection{Ultraviolet and Infrared fields}

We introduce a function
$\c_h(t,k_0) \in C^{\io}(\mathbb{T} \times  \mathbb{R})$,
such that $\c_h(t,k_0) = \c_h(-t,-k_0)$ and $\c_h(t,k_0) = 1$, if
$\sqrt{k_0^2+v^2 ||t||^2_1}\le a \g^{h-1}$
and  $\c_h(t,k_0) = 0$ if $\sqrt{k_0^2+v^2 ||t||^2_1}\le a \g^h$
with $a$ and $\g>1$ suitable constants. We choose $a$ so that
the supports of $\c_0(\o(x-\bar x),k_0)$ and $\c_0(\o(x+\bar x),k_0)$
are disjoint; note that the $C^\io$ function on $ \mathbb{T} \times  \mathbb{R}$
\be \hat \chi^{u.v.} (\o x,k_0) = 1- \c_0(\o(x-\bar x),k_0) -
\c_0(\o(x+\bar x),k_0)
\ee
is equal  to $0$, if $\sqrt{k_0^2+|\phi_x-\phi_{\bar x}|^2}\le b$, with $b$ a suitable constant.
For reasons which will appear clear below, we choose $\g> 2^{1\over\t}$. We  can write then
\be
g(\xx,\yy)=g^{(u.v)}(\xx,\yy)+g^{(i.r)}(\xx,\yy)
\ee
and
\be
g^{(i.r)}(\xx,\yy)=\sum_{\r=\pm} g_\r^{(\le 0)}(\xx,\yy)
\ee
where
\bea 
&&g^{(u.v.)}(\xx,\yy)={\d_{x,y}\over \b}\sum_{k_0\in
D_{\b}}\chi(\g^{-M}|k_0|) \hat \chi^{u.v.} (\o x,k_0)
{e^{-i k_0(x_0-y_0)}\over -i k_0+\phi_x-\phi_{\bar x}} \nn\\
&&g^{(\le 0)}_\r(\xx,\yy)={\d_{x,y}\over \b}\sum_{k_0\in
D_{\b}}\c_0(\o(x-\r\bar x),k_0) 
{e^{-i k_0(x_0-y_0)}\over -i k_0+\phi_x-\phi_{\bar x}} 
\eea
For definiteness, we start considering the generating function \pref{GI} with $\phi=0$.
The properties of Grassmann integrals imply that we can write 
\be e^{W(0)}=\int P(d\psi) e^{-\VV(\psi)} =
\int P(d\psi^{(i.r.)})\int P(d\psi^{(u.v.)}) e^{-\VV(\psi^{(i.r.)}+\psi^{(u.v.)})} \ee
where $P(d\psi^{(u.v.)})$ and $P(d\psi^{(i.r.)})$ are gaussian Grassmann integrations with propagators
respectively $g^{(u.v.)}(\xx,\yy)$ and $g^{(i.r)}(\xx,\yy)$ and $\psi^{(u.v.)}$ and $\psi^{(i.r.)}$ 
are independent Grassmann variables. We can write 
\be
\int P(d\psi^{(u.v.)}) e^{-\VV(\psi^{(i.r.)}+\psi^{(u.v.)})}=e^{\sum_{n=0}^\io{(-1)^n\over n!}\EE^T_{u.v.}(\VV:n)}\equiv e^{-\b L E_0-\VV^{(0)}(\psi^{i.r.})}
\ee
where $\EE^T_{u.v.}$ is the fermionic truncated expectation with respect to $P(d\psi^{(u.v.)})$; therefore
\be e^{W(0)}=e^{-\b L E_0}\int P(d\psi^{(i.r.)})e^{-\VV^{(0)}(\psi^{i.r.})} \ee
where 
\be
\VV^{(0)}=\sum_{n=2}^\io \sum_{x_1}\int dx_{0,1}....\sum_{x_n}\int dx_{0,n}
W_{n}^{(h)}
(\xx_1,...,\xx_n) [\prod_{i=1}^n \psi^{(\e_i)(\le 0)}_{\xx'_i+\r_i{\bar x}}]
\ee
Note that the kernel $W_{n}^{(h)}
(\xx_1,...,\xx_n)$ will contain in general Kronecker or Dirac
deltas, and we define the $L_1$ norm as they would be positive
functions. 

\begin{lemma}\lb{p2.2} The constant $E_{0}$ and the kernels
$W^{(0)}_{n}$ are given by power series in $\l,\e,\n$
convergent for $|\l|,|\e|, |\n|\le \e_0$, for $\e_0$ small enough and
independent of $\b,L,M$. They satisfy the following bounds:
\be |W^{(0)}_{n}|_{L_1}  \le \b L C^{n}
\e_0^{k_{n}}\;,\lb{2.17}\ee
for some constant $C>0$ and $k_{n}=\max\{1,n-1\}$. Moreover,
$\lim_{M\to\io}E_{0}$ and $\lim_{M\to\io} W^{(0)}_{n}$ do
exist and are reached uniformly, so that, in particular,
the limiting functions are analytic in the same domain.
\end{lemma}

\subsection{Proof of Lemma 3.1}

We can write
$\chi(\g^{-M}|k_0|)=\sum_{j=-\io}^M
f_j(|k_0|)$
with, for $j\le M-1$, 
$f_j(|k_0|)=\chi(\g^{-j}|k_0|)-\chi(\g^{-j+1}|k_0|)$ 
a smooth compact support function non vanishing for $\g^{h-1}\le |k_0|\le \g^{h+1}$.
. Therefore
\be g^{(u.v.)}(\xx,\yy)=\sum_{h=1}^{M} g^{(h)}(\xx,\yy)\;,\lb{B.1}\ee
where
\be
g^{(h)}(\xx,\yy)=\d_{x,y}{1\over\b}\sum_{k_0}{e^{i k_0 (x_0-y_0)}\over -i k_0+\phi_x-\phi_{\bar x }}
\chi^{(u.v.)}(k_0,\o x) f_h(|k_0|)=\d_{x,y} \bar g^{(h)}(x,x_0-y_0)
\label{B.2}
\ee
where we have used that $\chi(\g^{-N}|k_0|)=\sum_{h=1}^N f_h(|k_0|)$, according to the definition after 
\pref{limN}.
By integration by parts, for any integer $M$
\be
|\bar g^{(h)}(x,x_0-y_0)|\le {C_M\over 1+[\g^h|x_0-y_0|]^M}\label{xcx}
\ee
By using
\pref{B.1} we can write
$P(d\psi^{(u.v.)})=\prod_{h=1}^M P(d\psi^{(h)})$ and the
corresponding decomposition of the field $\psi^{(u.v.)}_{\xx,s}
=\sum_{h=1}^M \psi^{(h)}_{\xx,s}$. Hence, we can integrate
iteratively the fields $\psi^{(M)},\psi^{(M-1)},...,\psi^{(h)}$
with $h\ge 1$ and, if we define $\psi^{(\le 0)} = \psi^{i.r.}$ and
$\psi^{(\le h)} = \psi^{i.r.} + \sum_{j=1}^{h} \psi^{(j)}$, if
$h\ge 0$, we get :
\be\lb{2za} e^{\WW(0)} =e^{-L\b E_h}\; \int P(d\psi^{\le h}) \,
e^{-\VV^{(h)}(\psi^{(\le h)})} \ee
Let us consider first the effective potentials on scale $h$,
$\VV^{(h)}(\psi^{(\le h)})$. We want to show that they can be
expressed as sums of terms, each one associated to an element of a
family of labeled trees; we shall call this expansion {\em the
tree expansion}.
\vskip1cm
 The tree definition can be followed looking at
Fig 1.

\insertplot{700}{195}
{\ins{60pt}{90pt}{$v_0$}\ins{120pt}{100pt}{$v$}
\ins{100pt}{90pt}{$v'$}
\ins{120pt}{-5pt}{$h_v$}
\ins{235pt}{-5pt}{$M$}
\ins{255pt}{-5pt}{$M+1$}
}
{fig50}
{\label{h2q} 
A tree $\t\in\TT_{h,n}$ with its scale labels.
} {0}

Let us consider the family of all trees which can be constructed
by joining a point $r$, the {\it root}, with an ordered set of
$\bar n\ge 1$ points, the {\it endpoints} of the {\it unlabeled
tree}, so that $r$ is not a branching point. $\bar n$ will be
called the {\it order} of the unlabeled tree and the branching
points will be called the {\it non trivial vertices}. The
unlabeled trees are partially ordered from the root to the
endpoints in the natural way; we shall use the symbol $<$ to
denote the partial order. Two unlabeled trees are identified if
they can be superposed by a suitable continuous deformation, so
that the endpoints with the same index coincide. It is then easy
to see that the number of unlabeled trees with $\bar n$ end-points
is bounded by $4^{\bar n}$. We shall also consider the set
$\TT_{h,n,M}$ of the {\it labeled trees} with $n$ endpoints (to be
called simply trees in the following); they are defined by
associating some labels with the unlabeled trees, as explained in
the following items.

\0 2) We associate a label $h\le M$ with the root. Moreover, we
introduce a family of vertical lines, labeled by an integer taking
values in $[h,M+1]$, and we represent any tree $\t\in\TT_{M,h,n}$
so that, if $v$ is an endpoint or a non trivial vertex, it is
contained in a vertical line with index $h_v>h$, to be called the
{\it scale} of $v$, while the root $r$ is on the line with index
$h$. In general, the tree will intersect the vertical lines in set
of points different from the root, the endpoints and the branching
points; these points will be called {\it trivial vertices}. The
set of the {\it vertices} will be the union of the endpoints, of
the trivial vertices and of the non trivial vertices; note that
the root is not a vertex. Every vertex $v$ of a tree will be
associated to its scale label $h_v$, defined, as above, as the
label of the vertical line whom $v$ belongs to. Note that, if
$v_1$ and $v_2$ are two vertices and $v_1<v_2$, then
$h_{v_1}<h_{v_2}$.

\0 3) There is only one vertex immediately following the root, which will be
denoted $v_0$; its scale is $h+1$. If $v_0$ is an endpoint, the tree is
called the {\em trivial tree}; this can happen only if $n+m=1$.

\0 4) Given a vertex $v$ of $\t\in\TT_{M,h,n}$ that is not an
endpoint, we can consider the subtrees of $\t$ with root $v$,
which correspond to the connected components of the restriction of
$\t$ to the vertices $w\ge v$; the number of endpoint of these
subtrees will be called $n_v$. If a subtree with root $v$ contains
only $v$ and one endpoint on scale $h_v+1$, it will be called a
{\it trivial subtree}.

\0 5) Given an end-point, the vertex $v$ preceding it is surely a
non trivial vertex, if $n>1$.

\vspace{.3cm}

Our expansion is built by associating a value to any tree $\t\in
\TT_{M,h,n}$ in the following way.

First of all, given a normal endpoint $v\in\t$ with $h_v=M+1$, we
associate to it one of the terms (note that to the $\e$
interaction two terms are associated) contributing to the
potential $\VV^{(M)}(\psi)$ while, if
$h_v\le M$, we associate to it one of the terms appearing in the
following expression:
\be -\VV(\psi^{(< h_v)}) -\n \NN(\psi^{(< h_v)})+
\int d\xx (-\n_C(x+1)
+\l\bar g^{[h_v,M]}(x+1;0)) \psi^{+(<
h_v)}_{\xx}\psi^{-(< h_v)}_{\xx})\label{sssa}
\ee
$$+
\int d\xx (-\n_C(x) +\l \bar g^{[h_v,M]}(x;0)) \psi^{+(<
h_v)}_{\xx+{\bf e_1}}\psi^{-(< h_v)}_{\xx+{\bf e_1}}
$$
We associate to the label an index to specify which term is
associated to the end-point. We introduce also a {\it field label}
$f$ to distinguish the field variables appearing in the different
terms associated to the endpoints; the set of field labels
associated with the endpoint $v$ will be called $I_v$.
Analogously, if $v$ is not an endpoint, we shall call $I_v$ the
set of field labels associated with the endpoints following the
vertex $v$; $\xx(f)$, $\e(f)$  will denote the space-time point,
the $\e$ index of the Grassmann field variable with label $f$.

The previous definitions imply that, if $0\le h< M$, the following iterative
equations are satisfied:
\be -\VV^{(h)}(\psi^{(\le h)})-\b L e_h = \sum_{n=1}^\io
\sum_{\t\in\TT_{M,h,n}} \VV^{(h)}(\t,\psi^{(\le
h)})\;,\lb{B.12}\ee
where, if $v_0$ is the first vertex of $\t$ and $\t_1,\ldots,\t_s$, $s\ge 1$,
are the subtrees with root in $v_0$,
\be \VV^{(h)}(\t,\psi^{(\le h)})={(-1)^{s+1}\over s!} \EE^T_{h+1}
\big[\bar\VV^{(h+1)}(\t_1,\psi^{(\le h+1)});
\ldots;\bar\VV^{(h+1)} (\t_{s},\psi^{(\le
h+1)})\big]\;,\lb{B.13}\ee
where $\bar\VV^{(h+1)}(\t_i,\psi^{(\le h+1)})$ is equal to
$\VV^{(h+1)}(\t_i,\psi^{(\le h+1)})$ if the subtree $\t_i$
contains more than one end-point, otherwise it is given by one of
the terms contributing to the potentials in \pref{VM}, if
$h_v=M+1$, or one of the addends in \pref{sssa}, if $h_v\le M$,
the choice depending on the label $a$.

Note that
\be
|\n_C(x)|,|\l\bar g^{[h_v,M]}(x,0))|\le C|\l|\label{42}\ee
The above definitions imply, in particular, that, if $n>1$ and $v$
is not an endpoint, then $N_v>1$, with $N_v$ denoting the number
of endpoints following $v$ on $\t$; in fact the vertex preceding
an end-point is necessarily non trivial, if $n>1$.

Using its inductive definition, the right hand side of \pref{B.12}
can be further expanded, and in order to describe the resulting
expansion we need some more definitions.

We associate with any vertex $v$ of the tree a subset $P_v$ of
$I_v$, the {\it external fields} of $v$, and the set $\xx_v$ of
all space-time points associated with one of the end-points
following $v$. The subsets $P_v$ must satisfy various constraints.
First of all, $|P_v|\ge 2$, if $v>v_0$; moreover, if $v$ is not an
endpoint and $v_1,\ldots,v_{s_v}$ are the $s_v\ge 1$ vertices
immediately following it, then $P_v \subseteq \cup_i P_{v_i}$; if
$v$ is an endpoint, $P_v=I_v$. If $v$ is not an endpoint, we shall
denote by $Q_{v_i}$ the intersection of $P_v$ and $P_{v_i}$; this
definition implies that $P_v=\cup_i Q_{v_i}$. The union ${\cal
I}_v$ of the subsets $P_{v_i}\setminus Q_{v_i}$ is, by definition,
the set of the {\it internal fields} of $v$, and is non empty if
$s_v>1$. Given $\t\in\TT_{M,h,n}$, there are many possible choices
of the subsets $P_v$, $v\in\t$, compatible with all the
constraints. We shall denote ${\cal P}_\t$ the family of all these
choices and ${\bf P}$ the elements of ${\cal P}_\t$.

With these definitions, we can rewrite $\VV^{(h)}(\t,\psi^{(\le
h)})$ in the r.h.s. of \pref{B.12} as
\bea &&\VV^{(h)}(\t,\psi^{(\le h)})=\sum_{{\bf P}\in{\cal
P}_\t} \VV^{(h)}(\t,{\bf P})\;,\nn\\
&&\bar\VV^{(h)}(\t,{\bf P})=\int d\xx_{v_0} \widetilde\psi^{(\le
h)}(P_{v_0}) K_{\t,{\bf P}}^{(h+1)}(\xx_{v_0})\;,\lb{2.43a}\eea
where $K_{\t,{\bf P}}^{(h+1)}(\xx_{v_0})$ is defined inductively
by the equation, valid for any $v\in\t$ which is not an endpoint,
\be K_{\t,{\bf P}}^{(h_v)}(\xx_v)={1\over s_v !} \prod_{i=1}^{s_v}
[K^{(h_v+1)}_{v_i}(\xx_{v_i})]\; \;\EE^T_{h_v}[
\widetilde\psi^{(h_v)}(P_{v_1}\setminus Q_{v_1}),\ldots,
\widetilde\psi^{(h_v)}(P_{v_{s_v}}\setminus
Q_{v_{s_v}})]\;,\lb{2.45}\ee
Moreover, if $v_i$ is an endpoint, $K^{(h_v+1)}_{v_i}(\xx_{v_i})$
is equal to the kernel of one of the terms contributing to the
potential in \pref{VM}, if $h_{v_i}=N+1$, or one of the four terms
in \pref{sssa}, if $h_{v_i}\le N$; if $v_i$ is not an endpoint,
$K_{v_i}^{(h_v+1)}=K_{\t_i,{\bf P}_i}^{(h_v+1)}$, where ${\bf
P}_i=\{P_w, w\in\t_i\}$.

In order to get the final form of our expansion, we need a
convenient representation for the truncated expectation in the
r.h.s. of (\ref{2.45}). Let us put $s=s_v$, $P_i\=P_{v_i}\setminus
Q_{v_i}$; moreover we order in an arbitrary way the sets
$P_i^\pm\=\{f\in P_i,\e(f)=\pm\}$, we call $f_{ij}^\pm$ their
elements and we define $\xx^{(i)}=\cup_{f\in P_i^-}\xx(f)$,
$\yy^{(i)}=\cup_{f\in P_i^+}\yy(f)$, $\xx_{ij}=\xx(f^-_{ij})$,
$\yy_{ij}=\xx(f^+_{ij})$. Note that $\sum_{i=1}^s
|P_i^-|=\sum_{i=1}^s |P_i^+|\=k$, otherwise the truncated
expectation vanishes. A couple
$l\=(f^-_{ij},f^+_{i'j'})\=(f^-_l,f^+_l)$ will be called a line
joining the fields with labels $f^-_{ij},f^+_{i'j'}$. Then, we use
the {\it Brydges-Battle-Federbush} formula,  if $s>1$,
\be \EE^T_{h}(\widetilde\psi^{(h)}(P_1),\ldots,
\widetilde\psi^{(h)}(P_s))=\sum_{T}\prod_{l\in T}
\big[g^{(h)}(\xx_l-\yy_l)\big]\, \int dP_{T}({\bf t})\; {\rm
det}\, G^{h,T}({\bf t})\;,\lb{2.46a}\ee
where $T$ is a set of lines forming an {\it anchored tree graph} between the
clusters of points $\xx^{(i)}\cup\yy^{(i)}$, that is $T$ is a set of lines,
which becomes a tree graph if one identifies all the points in the same
cluster. Moreover ${\bf t}=\{t_{ii'}\in [0,1], 1\le i,i' \le s\}$,
$dP_{T}({\bf t})$ is a probability measure with support on a set of ${\bf t}$
such that $t_{ii'}={\bf u}_i\cdot{\bf u}_{i'}$ for some family of vectors
${\bf u}_i\in \RRR^s$ of unit norm.
\be G^{h,T}_{ij,i'j'}=t_{ii'}
\d_{x_{ij},y_{i'j'}}
\big[\tilde g^{(h)}(x_{ij}, x_{0,ij}-y_{0,i'j'})\big]_{\r^-_{ij},\r^+_{i'j'}}\;,
\label{2.48}\ee
with $(f^-_{ij}, f^+_{i'j'})$ not belonging to $T$.

\insertplot{930}{195}
{}
{fig2}
{\label{h2q} 
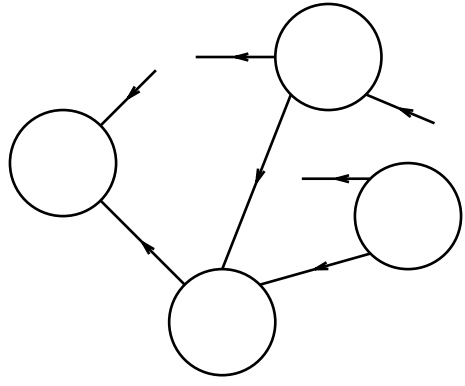
A tree $T_v$ connecting $S_v$ terms
; inside the circles are the trees $T_{\bar v}$ with $v<\bar v$} 
{0}

By inserting \pref{2.46a} in the r.h.s. of \pref{2.45} we get
\be V^{(h)}(\t,{\bf P} )=\sum_{T\in {\bf T}} \int d\xx_{v_0}
W_{\t,{\bf P},T}(\xx_{v_0})\prod_{f\in P_{v_0}}\psi^{(\le
h)\s(f)}_{\xx(f)} \ee where
\be W_{\t,{\bf P},T}(\xx_{v_0})=\prod_{v\,\hbox{\rm not
e.p.}}{1\over s_v!} \int dP_{T_v}({\bf t}_v) det G^{h_v,T_v}({\bf
t}_v) \prod_{l\in T_v}\d_{x_\ell,y_\ell } \bar
g^{(h_v)}(x_\ell;x_{0,\ell}-y_{0,\ell}l)\ee
${\bf T}$ is the set of the tree graphs on $\xx_{v_0}$, obtained
by putting together an anchored tree graph $T_v$ for each non
trivial vertex $v$; $v^*_1,\ldots,v^*_n$ are the endpoints of
$\t$, $f^-_l$ and $f^+_l$ are the labels of the two fields forming
the line $l$, ``e.p.''  is an abbreviation of ``endpoint''.

Note that we can eliminate 
the Kronecker deltas in the propagators in the spanning tree $T$, so that only a single sum over the coordinate remain and the coordinate of the external fields 
and of the fields in the determinants are assigned once that $x$,  $T$
and $\t$ are given, as the interaction is quasi local;
we can then write 
\be V^{(h)}(\t,{\bf P} )=\sum_{T\in {\bf T}} \sum_x \int
dx_{0,v_0} H_{\t,{\bf P},T}(x, x_{0,v_0})\prod_{f\in
P_{v_0}}\psi^{(\le h)\s(f)}_{\hat \xx(f)} \label{zac} \ee
where
\be H_{\t,{\bf P},T}(x, x_{0,v_0})= \prod_{v\,\hbox{\rm not
e.p.}}{1\over s_v!} \int dP_{T_v}({\bf t}_v) det G^{h_v,T_v}({\bf
t}_v) \prod_{l\in T_v} \bar g^{(h_v)}(\hat
x_\ell; x_{0,\ell}-y_{0,\ell}) ] \ee
where there is a field $\bar f$ such that
$\hat x(\bar f)=x$ and all the other coordinates $\hat x(f)$ are assigned
once that $x$, $T$ and $\t$ are given. We will call {\it resonances} the terms such
that $\hat \xx(f)=x$ for ant $f\in P_{v_0}$.

In order to bound the above expression we introduce an Hilbert
space $\HH=\RRR^L\otimes\RRR^s\otimes L^2(\RRR^1)$ so that
\be
G^{h,T}_{ij,i'j'}=
\Big({\bf v}_{x_{ij}}\otimes {\bf u}_{i}\otimes
A(x_{0, ij}-,x_{ij})\;,\ {\bf v}_{y_{i',j'}}\otimes
{\bf u}_{i'}
\otimes B(y_{0,i'j'}-,x_{ij})\Big)
\label{as}
\;,\ee
where ${\bf v}\in \RRR^{L}$ are unit vectors such that $({\bf v}_i,{\bf v}_j)=\d_{ij}$,
${\bf u}\in \RRR^{s}$ are unit vectors $(u_i,u_{i})=t_{ii'}$, and $A,B$
are vectors in the Hilbert space with scalar product \be (A,B)=\int dz_0 A(x',x_0-z_0)B^*(x',z_0-y_0)\ee
given by
\bea &&A(x,x_0-z_0)={1\over\b}\sum_{k_0} e^{-i k_0
(x_0-z_0)}\sqrt{\chi^{(u.v.)} f_h(|k_0|)}(k_0^2+(\phi_x-\m)^2)
\nn\\
&&B(x,y_0-z_0)={1\over \b}\sum_{k_0} e^{-ik_0( y_0-z_0)}
\sqrt{\chi^{(u.v.)} f_h(|k_0|)}(-i k_0+\phi_x-\m)\;. \label{2.48b}\eea
Moreover
\be ||A_h||^2=\int dz_0 |A_h(\zz)|^2\le C\g^{-3h}\;,\quad\quad
||B_h||^2\le C \g^{3h}\;,\lb{B.5}\ee
for a suitable constant $C$.

If $\e_0=\max \{|\l|,|\n|\}$, by using \pref{2.45} and \pref{2.46a}, we get the bound
\bea &&{1\over \b L}\sum_{\t\in {\cal T}_{M,h,n}} \sum_{T\in {\bf
T}}\sum_{{\bf P}\in \PP_\t} \sum_x \int dx_{0,v_0} |H_{\t,{\bf
P},T}(x, x_{0,v_0})| \le\\
&& \sum_{\t\in {\cal T}_{M,h,n}} \sum_{T\in {\bf T}}\sum_{{\bf
P}\in \PP_\t} \Bigg[\prod_{v\ {\rm not}\ {\rm e.p.}}{1\over s_v!}
\max_{{\bf t}_v}\big|{\rm det}\, G^{h_v,T_v}({\bf t}_v)\big|
\prod_{l\in T_v} \prod_{l\in T_v} \int d(x_{0,l}-y_{0,l}) |\sup_x
|\bar g^{(h_v)}(x_l;x_{0,l}-y_{0,l})|| \big|\Bigg]\nn \eea
where, given the tree $\t$, $\bf T$ is the family of all tree graphs joining the
space-time points associated to the endpoints, which are obtained by taking,
for each non trivial vertex $v$, one of the anchored tree graph $T_v$ appearing
in \pref{2.46a}, and by adding the lines connecting the two vertices
associated to non local endpoints. Note that the sum over the spatial coordinates is trivial
thanks to the $\d_{x,y}$ present in the propagators. Gram--Hadamard inequality, combined with
\pref{B.5}, implies the dimensional
bound:
\be |{\rm det} G^{h_v,T_v}({\bf t}_v)| \le
C^{\sum_{i=1}^{s_v}|P_{v_i}|-|P_v|-2(s_v-1)}\;.\lb{2.54a}\ee

By the decay properties of $g^{(h)}(\xx)$ given by \pref{xcx}, it
also follows that
\be\lb{2.55a} \prod_{v\ {\rm not}\ {\rm e.p.}} {1\over s_v!} \prod_{l\in
T_v} \int d(x_{0,l}-y_{0,l})
|\sup_x |\bar g^{(h_v)}(x_l;x_{0,l}-y_{0,l})||\le C^{n+m} \prod_{v\
{\rm not}\ {\rm e.p.}} {1\over s_v!} \g^{-h_v(s_v-1)}\ee
We can now perform the sum $\sum_{T\in{\bf T}}$, which erases the $1/s_v!$ up
to a $C^n$ factor. Then, by using the identity $\sum_{v'\ge v} (s_{v'}-1) =
n_v-1$ and the bound $\sum_{v\ge v_0}
[\sum_{i=1}^{s_v}|P_{v_i}|-|P_v|-2(s_v-1)] \le 4n-2(n-1)$, we easily get
the final bound
\be \sum_{n=1}^\io C^{n} \e_0^n \sum_{\t\in {\cal
T}_{M,h,n}} \sum_{{\bf P}  \in{\cal P}_\t\atop |P_{v_0}|=0}
\g^{-h(n-1)}\Big[\prod_{v\ \text{not trivial}}
\g^{-(h_v-h_{v'})(N_v-1)}\Big]\ee
where $v'$ is the non trivial vertex immediately preceding $v$ or $v_0$. This
bound is suitable to control the expansion, if $n>1$, since $N_v>1$ for any
non trivial vertex, as discussed below \pref{42}. If $n=1$ the allowed
trees have only one endpoint of scale $h+1$.

Note that $\sum_{T\in{\bf T}}$ can be bounded by $\prod_v s_v!
C^{\sum_{i=1}^{s_v}|P_{v_i}|-|P_v|-2(s_v-1)} \le c^{n}\prod_v s_v!$.
In order to bound the sum over $\t$, note that the number of unlabeled trees is $\le 4^n$; moreover, as $N_v>1$ and, if $v>v_0$,
$2\le |P_v|\le 4N_v -2(N_v-1)$, so that $N_v-1 \ge |P_v|/6$,
\be \Big[\prod_{v\ \text{not trivial}} \g^{-(h_v-h_{v'})(N_v-1)}\Big]\le
\Big[\prod_{v\ \text{not trivial}} \g^{- {2\over 5}(h_v-h_{v'})}\Big]
\Big[\prod_{v\ {\rm not}\ {\rm e.p.}}\g^{- {|P_v|\over 10}}\Big]\lb{B.18b}
\ee
The factor $\g^{- {2\over 5}(h_v-h_{v'})}$ can be used to bound the sum over
the scale labels of the tree; moreover
\be \sum_{\bP\in{\cal P}_\t} \g^{- {|P_v|\over 10}}\le C^{n}\lb{B.18c}\ee
Since the constant $C$ is independent of $M,\b,L$, the bounds above imply
analyticity of the kernels in $\l$ and $\n$, if $\e_0$ is small enough. It is an immediate consequence of the above bounds the proof of 
uniform convergence of the $M\to\io$ limit; the proof of this is essentially identical to the one in
\cite{BFM} after (2.8) and it will not repeated here.
\qed

\section{The infrared integration and the small divisor problem}

\subsection{Multiscale analysis}
 
The integration of the infrared (negative)  scales has to be done in  a different way, including the quadratic terms present in the effective potential producing a mass term.
We describe the integration of the infrared scales by iteration; assume that we have integrated the fields $\psi^{(0)}...\psi^{(h)}$
obtaining
\be  e^{-\b L E_0}\int P(d\psi^{(\le 0)})e^{\VV^{(0)}(\psi^{(\le 0)})}=e^{-\b L E_h} \int P(d\psi^{(\le h)})e^{-\VV^{(h)}(\psi^{(\le h)})}\label{gr1} \ee
where $P(d\psi^{(\le h)})$ is the gaussian grassman measure with propagator, $\r=\pm $
\be
g^{(\le h)}_{\r,\r'}(\xx',\yy')=\d_{x',y'}\bar g^{(\le h)}_{\r.\r'}(x',x_0-y_0)
\ee
with
\bea
&& g^{(\le h)}_{\r,\r'}(x',x_0-y_0')= \nn\\
&&= \int dk_0 e^{i k_0 (x_0-y_0)}\chi_h(\o x',k_0) 
\begin{pmatrix}
&-i
k_0+ v \o x'+r_{x'} & \s_h\\ &\s_h & -i k_0- v \o x'+r_{x'}
\end{pmatrix}^{-1}_{\r,\r'}\equiv\nn\\
 &&\int dk_0 e^{i k_0 (x_0-y_0)}\chi_h(\o x',k_0)
A^{-1}_{h,\r,\r'}(x',k_0)
\label{prop1}
\eea
where
%
$\VV^{(h)}$ can be written as sum over trees (similar to the ones
for $\VV^{(0)}$ and defined precisely below), and each tree with $n$ end points
contribute to $\VV^{(h)}$ with a term of the form, after integrating the Koenecker deltas
in the spanning tree as discussed before \pref{zac}
\be \sum_{x'}\int dx_{0,1}....\int dx_{0,n}
H_{n;\r_1,..,\r_n}^{(h)} (x'_1; x_{0,1},.,x_{0,n}) [\prod_{i=1}^n
\psi^{(\e_i)(\le h)}_{\xx'_i, \r_i}] \ee
where the coordinates of the external fields $x'_i$ are 
assigned once that $x$ and the labels of the tree are
assigned. As in
the previous section we call resonances the terms such that all
the coordinates of the external points are equal
\be x'_i=x'_1 \label{gr}\ee
We can split $\VV^{(h)}$ in two parts
\be
\VV^{(h)}=\VV^{(h)}_R+\VV^{(h)}_{NR}
\ee
where in $\VV^{(h)}_R$ are the resonant term (whose external
fields verify \pref{gr}) while $\VV^{(h)}_{NR}$ are the remaining
terms.

We define a {\it localization} operation as a linear operation
acting on $\VV^{(h)}$ in the following way:
\begin{enumerate}
\item On the non resonant part of the effective potential is
defined as \be \LL\VV^{(h)}_{NR}=0\ee \item On the resonant part
of the effective potential its action consists in setting the time coordinate of the external
fields equal \bea &&\LL \sum_{x'} \int dx_{0,1},,,x_{0,n}
H_{n,\r_1,..,\r_n}^{(h)}(x';x_{0,1},..,x_{0,n})[\prod_{i=1}^n
\psi^{(\e_i)(\le h)}
_{x',x_{0,i},\r_i}]=\nn\\
&& \sum_{x'} \int dx_{0,1},,,x_{0,n}
H_{n,\r_1,..,\r_n}^{(h)}(x';x_{0,1},..,x_{0,n})[\prod_{i=1}^n
\psi^{(\e_i)(\le h)} _{x',x_{0,1},\r_i}] \eea
\end{enumerate}
We can write
\be \LL \VV^{(h)}= \g^h \n_h F^{(h)}_\n+  F^{(h)}_{z}
+s_h F^{(h)}_\s+\ F^{(h)}_{\z}
+F_\l^{(h)} =s_h F^{(h)}_\s+\bar\LL\VV^{(h)} \ee
where (note that $H^h_n$ is translation invariant in the time
direction)
\bea
&&s_h={1\over \b}\int dx_0  dy_0H^{(h)}_{2,\r-\r}(0,x_0,y_0)\quad\quad \z_h(x')=
{1\over \b}\int dx_0  dy_0 \partial_x H^{(h)}_{2,\r-\r}(x',x_0,y_0)\\
&&
\n_h={1\over\b}\int dx_0  dy_0 H^{(h)}_{2,\r\r}(x',x_0,y_0)\quad\quad z_h(x')={1\over\b}\int dx_0  dy_0 \partial_x H^{(h)}_{2,\r\r}(x',x_0,y_0)\nn\\
&&
\l_{h}(x')={1\over\b}\int d x_{0,1}...dx_{0,4} H^{(h)}_{4}(x';x_{0,1},x_{0,2},x_{0,3},x_{0,4})\eea
and 
\bea &&F^{(h)}_\n= \sum_{\r}\sum_{x'}\int dx_0 \hat\psi^{+(\le
h)}_{\xx',\r}
\hat\psi^{-(\le h)}_{\xx',\r}\quad\quad F^{(h)}_\z= \sum_{\r}\sum_{x'} \int dx_0 (\o x') \z_h(x')\psi^{+(\le
h)}_{\xx',\r}
\hat\psi^{-(\le h)}_{\xx',-\r}\nn\\
&&F^{(h)}_\s= \sum_{\r}\sum_{x'}\int dx_0 \hat\psi^{+(\le
h)}_{\xx',\r}
\hat\psi^{-(\le h)}_{\xx',-\r}\quad F^{(h)}_z= \sum_{\r}\sum_{x'}\int dx_0 (\o x')
z_h(x')
\psi^{+(\le
h)}_{\xx',\r}
\hat\psi^{-(\le h)}_{\xx',\r}\nn\\
&&F^{(h)}_\l=\sum_{x'}\int dx_0 \l_h(x')\psi^{+(\le h)}_{\xx',+}
\psi^{-(\le h)}_{\xx',+} \psi^{+(\le h)}_{\xx,-}
\psi^{-(\le h)}_{\xx',-} \eea
{\it Note that the local terms with more than 6 fields  are vanishing for
$n\ge 6$} as there are at least two fields with the same
$\e,\r$ and the same coordinate. Therefore the
$\LL$ operation produces non vanishing terms only on the tems with
$n=2,4$. Note that $\s_0=O(\e^{2\bar x})$.

We also define a {\it renormalization} operation as \be
\RR=1-\LL\ee so that we can rewrite \pref{gr1} as
\bea 
&&\int P(d\psi^{(\le h)})e^{-\LL\VV^{(h)}(\psi)-\RR
\VV^{(h)}(\psi)}=
 \int P(d\psi^{(\le h)})e^{-s_h F^{(h)}_\s-\bar\LL\VV^{(h)}
-\RR\VV^{(h)}}=
\label{gr1} \\
&& \int \tilde P(d\psi^{(\le h)})e^{-\bar\LL \VV^{(h)}
-\RR \VV^{(h)}}\nn
\eea
with $\tilde P(d\psi^{(\le h)})$ with a propagator $\tilde g^{(\le h)}$ coinciding with $g^{(\le h)}$ with $\s_h$
replaced by $\s_{h-1}$ with
\be
\s_{h-1}=\s_h+\chi_h s_h
\ee
The effect of the $\RR$ operation is the following
\bea &&\RR\sum_{x'}\int dx_{0,1}....\int dx_{0,n}
H_{n;\r_1,..,\r_n}^{(h)} (x'; x_{0,1},.,x_{0,n}) [\prod_{i=1}^n
\psi^{(\e_i)(\le h)}_{x',x_{0,i} ,\r_i}]=\sum_{x'}\int
dx_{0,1}....\int dx_{0,n}\label{r1}\nn
\\
&&\{ H_{n,\r_1,..,\r_n}^{(h)}(x';x_{0,1},..,x_{0,n})[\prod_{i=1}^n
\psi^{(\e_i)(\le h)}
_{x',x_{0,i},\r_i}-\prod_{i=1}^n
\psi^{(\e_i)(\le h)} _{x',x_{0,1},\r_i}]\}\eea
%

We write then
\be
\int  P(d\psi^{\le h-1})\int P(d\psi^{(h)}
)e^{-\LL\bar\VV^{(h)}-\RR\VV^{(h)}}=e^{-\b L \tilde E_h}
\int P(d\psi^{(\le h-1)}) e^{-\VV^{(h-1)}(\psi^{(\le h-1)})}\label{sss}
\ee
where $P(d\psi^{\le h-1})$ have
propagator $g^{(\le h-1)}$ coinciding with 
\pref{prop1} with $h-1$ replacing $h$, and 
$P(d\psi^{(h)}$ has propagator $g^{(h)}$ coinciding with with 
 $g^{(\le h-1)}$ with $\chi_{h-1}$ replaced by $f_h=\chi_{h-1}-\chi_h$,
where $f_h$ a smooth has support in $c\g^{h-1}\le |k_0^2+v_0^2 ||\o x'||_1^2\le c\g^{h+1}$, for a suitable constant $c$.
%
%
Starting from the r.h.s. of \pref{sss}, the procedure can be iterated.
Note that, for any integer $N$  and a
suitable constant $C_N$
\be
|\bar g^{(h)}(x,x_0-y_0)| \le {C_N\over 1+(\g^h|x_0-y_0|)^N}\label{za}
\ee
The above bound can be easily obtained integrating by parts.

\subsection{Tree expansion}

Again $\VV^{(h)}$ can be written as sum over trees, up to the following modifications to take into account the different multiscale integration procedure.
\begin{enumerate}
\item
The scale index now is an integer taking values in $[h,2]$, $h$ being
the scale of the root.
\item
With each vertex $v$ of scale $h_v=+1$, which is not an endpoint, we
associate one of the terms contributing to $-\VV^{(0)}(\psi^{(\le
0)})$, in the limit $M=\io$. With 
each endpoint $v$ of scale $h_v\le 1$ we associate one of local
terms that contribute to $\LL\VV^{(h_v-1)}$, and there is the constrain that
$h_v=h_{v'}+1$, if $v'$ is the non trivial vertex immediately preceding
it or $v_0$; to the end-points of scale $h_v=2$ are
associated one of the terms
contributing to  $-\VV$ and there is not such a constrain.
\item
 With each trivial or non trivial vertex $v>v_0$, which is not an endpoint, we
associate the $\RR=1-\LL$ operator, acting on the corresponding kernel.
\end{enumerate}

\insertplot{600}{195}
{\ins{50pt}{80pt}{$v_0$}
\ins{50pt}{-5pt}{$h$}
\ins{95pt}{-5pt}{$h_{v'}$}
\ins{120pt}{-5pt}{$h_{v}$}
\ins{100pt}{90pt}{$v'$}
\ins{115pt}{95pt}{$v$}
\ins{220pt}{-5pt}{$0$}\ins{240pt}{-5pt}{$1$}\ins{260pt}{-5pt}{$2$}}
{fig51}
{\label{h2q} 
A tree $\t\in\TT_{h,n}$ with its scale labels.}
{0}

A vertex $v$ which is not an end-point such that the spatial coordinates $x'$ in $P_v$ are all equal is called {\it resonant vertex},
while if the coordinates are different is called {\it non resonant vertex}; the set of resonantl vertices is denoted by$ H$ and the set on non-resonant vertices is denoted by $L$.
If $v_1,\ldots,v_{S_v}$
are the $S_v\ge 1$ vertices following the vertex $v$, we define
\be
S_v=S^L_v+S^H_v+S^2_v\label{bra}
\ee
where $S^L_v$ is the number of {\it non resonant} vertices following $v$,
$S^H_v$  is the number of {\it resonant} vertices following $v$,
while $S_v^2$  is the number of trivial trees with root $v$ associated to end-points.

If $h\le -1$,
the effective potential can be written in the following way:
\be
\VV^{(h)}(\psi^{(\le h)}) + L\b \tilde E_{h+1}=
\sum_{n=1}^\io\sum_{\t\in\TT_{h,n}}
V^{(h)}(\t,\psi^{(\le h)})
\ee
where, if $v_0$ is the first vertex of $\t$ and $\t_1,..,\t_s$ ($s=s_{v_0}$)
are the subtrees of $\t$ with root $v_0$,\\
$V^{(h)}(\t,\psi^{(\le h)})$ is defined inductively by the relation, if $s>1$
\be
V^{(h)}(\t,\psi^{(\le h)})=
{(-1)^{s+1}\over s!} \EE^T_{h+1}[\bar
V^{(h+1)}(\t_1,\psi^{(\le h+1)});..; \bar
V^{(h+1)}(\t_{s},\psi^{(\le h+1)})]\label{3.33}
\ee
where $\bar V^{(h+1)}(\t_i,\psi^{(\le h+1)})$:
\begin{enumerate}
\item 
it is equal to $\RR\VV^{(h+1)}(\t_i,\psi^{(\le h+1)})$, with $\RR$ given by \pref{r1}, if
the subtree $\t_i$ is non trivial;
\item if $\t_i$ is trivial and $h\le -1$, it is equal to one of the terms of $\LL\VV^{h+1}$ or, if $h = 0$, to one of the terms in the
$\VV$.
\end{enumerate}
By using \pref{3.33} and the representation of the truncated
expectations we get

\be V^{(h)}=\sum_{n=1}^\io \sum_{\t\in \TT_{h,n}}  \sum_{T\in {\bf
T}} \sum_x \int dx_{0,v_0} H_{\t,{\bf P},T}(x,
x_{0,v_0})\prod_{f\in P_{v_0}}\psi^{(\le h)\s(f)}_{\hat
\xx(f)}\Big\} \label{lau}\ee
where one of the spatial coordinates of the external fields
$\prod_{f\in P_{v_0}}\psi^{(\le h)\s(f)}_{\xx(f)}$ is equal
to $x'$ and the others are determined according to the 
following rule.
\begin{enumerate}
\item
We define a tree $\bar T_v$ starting
from $T_v$ and attaching to it the trees $T_{v_1},..,T_{v_{S_v}}$ associated to the vertices $v_1,..,v_{S_v}$
following $v$ (graphically, we consider the tree $T_v$ in Fig. 2 and we replace the bubbles with the corresponding subtrees), and repeating
this operation until the end-points are reached.
The tree $\bar T_v$ is composed by 
a set of lines, representing propagators with scale $h_{\bar v}\ge h_v$, connecting end-points $w$ of the tree $\t$. Note that, contrary to $T_v$, the vertices of $\bar T_v$ are connected with at most four lines.
\item To each vertex $w$ of $\bar T_v$ is associated a coordinate $x_w$; if there are external fields $\psi^{(\le h_v)}$ with coordinate $x_w$, we represent them as wiggly lines (see Fig. 4).
\item
To each line $\ell$ of $\bar T_v$ we associate a label $a_{\ell}=0,\pm 2\bar x$ 
\item
To each vertex $w$ of $\bar T_v$ is associated a coordinate $x_w$ and to each
line coming in or out $w$ is associated a factor $\d_{w}^{i_w}$, where $i_w$
is a label identifying the lines connected to $w$. The vertices $w$ (which corresponds to the end-points of $\t$) can be of type $\l,\n$ or $\l_h,z_h,\z_h$, and: 
\item $\d^i_w=0$ if $w$ if it corresponds to a $\n$ or $\n_h,z_h$ end-point;  $\d^i_w=\pm 2\bar x$ if $w$ if it correspond to a $\tilde\z_h$ end-point;
\item $\d_w^i=\pm 1$ it corresponds to an $\e$ end-point;
$\d^i_w=(0, \pm 1)$ is a $\l$ end-point; $\d^i_w=(0,\pm 2 \bar x)$
if is a $\l_h$ end-point
\end{enumerate}

\insertplot{800}{195}
{\ins{100pt}{110pt}{$w_1$}
\ins{75pt}{100pt}{$w_a$}
\ins{45pt}{80pt}{$w_b$}
\ins{20pt}{80pt}{$w_c$}
\ins{40pt}{40pt}{$w_2$}
}
{fig60}
{\label{h2} A tree $\bar T_v$
} {0}

According to the above definitions, consider two
vertices $w_1,w_2$ such that $x'_{w_1}$ and $x'_{w_2}$ are
coordinates of the external fields, and  let be $c_{w_1,w_2}$ the path (vertices and lines)
in $\bar T_v$ connecting $w_1$ with $w_2$ (in the example in Fig. 4 the path is composed by $w_1,w_a,w_b,w_c,w_2$ and the corresponding lines)
; as the path is a linear tree there is a natural orientation in the vertices, and we we call $i_w$ the label of the line exiting fom $w$ in 
$c_{w_1,w_2}$.
Therefore the following relation holds
\be x'_{w_1}-x'_{w_2}= (\r_{\ell_{w_2}}-\r_{\ell_{w_1}})\bar x+ \sum_{w\in c_{w_1,w_2}}
\d_{w}^{i_{w}}+\sum_{\ell\in c_{w_1,w_2}}a_\ell\label{fa}
\ee

The Diophantine condition implies a relation between the
scale $h_v$ and the number of vertices between 
$w_2$ and
$w_1$.

\begin{lemma}
If $|c_{w_1,w_2}|$ is the number of vertices in the path 
$c_{w_1,w_2}$ 
with $x'_{w_1}\not=x'_{w_2}$
than, if $v'$ is the first vertex following $v$ in $\t$
\be |c_{w_1,w_2}|\ge A \bar x^{-1} \g^{-h_{ v'}\over \t}\label{h19}
\ee with a suitable constant $A$.
\end{lemma}

{\it Proof.} Note that $|\o x'_{w}|\le c v_0^{-1} \g^{h_{v'}-1}$; there by
using \pref{fa} and the Diophanine condition
\bea
&&2 c v_0^{-1}\g^{h_{v'}}\ge ||\o x'_{w_1}||_1+||\o x'_{w_2}||_1 \ge ||\o(x'_{w_1}-x'_{w_2})
||_1\\
&&=||\o(
 (\r_{\ell_{w_2}}-\r_{\ell_{w_1}})\bar x+ \sum_{w\in c_{w_1,w_2}}
\d_{w}^{i_{w}}+\sum_{\ell\in c_{w_1,w_2}}a_\ell
)
||_1\ge \nn\\
&&C_0 | (\r_{\ell_{w_2}}-\r_{\ell_{w_1}})\bar x+ \sum_{w\in c_{w_1,w_2}}
\d_{w}^{i_{w}}+\sum_{\ell\in c_{w_1,w_2}}a_\ell
|^{-\t} \ge C_0 (4
\bar x |c_{w_2,w_1}|)^{-\t}\nn \eea
\qed \vskip.3cm 
The relation \pref{fa} is the
analogue of the conservation of momentum rule in ordinary Feynman
graphs. Lemma 3.1 says that there is a relation between the number
of vertices and the scale of the external lines in the non
resonant vertices; it is the analogue of Bruno lemma in KAM
theory.

\subsection{Power counting improvement and Diophantine condition}

We define $v_h=\e\tilde v_h$
where $v_h$ are the running coupling constants.
Therefore, each contribution from the tree $\t\in\TT_{h,n}$ is
proportional to a factor $\e^n$.
\begin{lemma}
If $v$ is a vertex of $\t$ in $\t\in \TT_{h,n}$ which is not end-point, and $v^*_i$ the end-points, and $N_v=\sum_{i, v^*_i>v} 1$
the number of end-points following $v$ then
\be
 \e^{n} \le \e^{n\over 2}\prod_{v not e.p.} \e^{N_v 2^{h_{ v'}-1}}\label{org}
\ee
where $v'$ is the vertex following $v$ in $\t$.
\end{lemma}
\vskip.3cm {\it Proof} We can write
\be
\e^{1\over 2}=\prod_{h=-\io}^0 \e^{2^{h-2}}
\ee
Given a tree $\t\in \TT_{h,n}$, we consider an end-point $v^*$ and the path in $\t$ from $v^*$
to the root $v_0$; to each vertex $v$ in such path with scale $h_v$ we associate a factor 
$\e^{2^{h_v-2}}$; repeating such operation for any end-point, the vertices $v$ followed by $N_v$ end-points are in $N_v$ paths, therefore  we can associate to them a factor  $\e^{N_v 2^{h_{ v}-2}}$; finally
we use that $h_{v'}=h_v-1$.
\qed 
\vskip.3cm It is an
immediate consequence of Lemma 3.1 and Lemma 3.2 the following
result, ensuring that we can extract from the $\e^n$ factor a
small factor to be associated to the non resonant vertices.
\begin{lemma}
For any tree $\t$, if $L$ is the set of non resonant vertices  
\be \e^{n\over 4}\le \prod_{v\in L} \e^{ A \bar
x^{-1} \g^{-h_{ v'}\over \t}2^{h_{ v'}-1}} \label{zza} \ee
\end{lemma}
\vskip.3cm
{\it Proof.}
Note that if $v$ is non resonant, there exists surely two external fields
with coordinates $x'_1, x'_2$ such that $x'_1\not=x'_2$; note that
\be
N_v\ge |c_{w_1,w_2}|\ge A \bar x^{-1} \g^{-h_{v'}\over \t}
\ee
therefore
\be
\e^{n\over 4}\le   \prod_{v\in L}\e^{  A \bar x^{-1} \g^{-h_{v'}\over \t} 2^{-h_{ v'}+1}}
\ee
%
\qed.
\vskip.6cm

\subsection{Renormalized expansion}

We want now to write the effective potential $\VV^{(h)}$ , see \pref{3.33}, in an equivalent way which is more suitable for the final bounds. 
Given a contribution
to $V^h$ corresponding to $\t,T,P$, we consider the vertex $v^*$ of $\t$
which are either non trivial or trivial but with some self-contraction of the external fields.
We call vertices $v^*$ of depth $k$ the $v^*$ followed vertices of depth $<k$ and at least one with depth $k-1$; the vertices
$v^*$ followed only by end-points are of depth $1$

We start from the vertices $v^*$ of depth $1$
and we distinguish three cases : 1) $v$ is non
resonant or $|P_v|\le 4$; 2) $v$ is resonant 3) $|P_v|\ge 6$. 
In case 2) the $\RR$ operation
acts on the external fields, as in  \pref{r1}; we can rewrite the difference of fields in the r.h.s.
of \pref{r1} as sum of terms in which a field $\psi$ is replaced by a field $D$ defined as
\be  D^{\e(\le h_v)}_{x_{1,0},x_{2,0} x',\r}= \psi^{(\e)(\le h_v)}_{x_{1,0},x',\r}- \psi^{(\e)(\le h_v)}_{x_{2,0},x',\r}\ee The corresponding
propagator can be written as
\be g^{(h_{v'})}(x_{1,0}-z_0,x')-g^{(h_{v'})}(x_{2,0}-z_0,\bar
x')=(x_0-y_0) \int_0^1 dt \partial  g^{(h_{v'})}(t
(\hat x_{12,0}(t)-z_0,x')
\label{df11} \ee
where $\hat x_{12,0}(t)=x_{1,0}+t(x_{2,0}-x_{1,0})$ is an interpolated point between $x_{0,1}$
and $x_{0,2}$.
We introduce then an extra label to identify in the expansion the external fields of $v$ to which is associated a $\psi$ or $D$ field. 

Let us consider now case 3).  
If
$|P_v|\ge 6$ we call $\bar\r,\bar\e$ the labels of the external field whose number is maximal; the number of such external lines is $\ge {|P_v|/4}$.
We consider a tree $\bar T_v$ and we associate to it another tree $\hat T_v$
eliminating from $\bar T_v$ all the trivial vertices not associated 
to any external line
with label $\bar \r,\bar\e$, and all the 
subtrees not containing any external line
with label $\bar \r,\bar\e$ (see Fig. 5 for an example), so that there is at least an external line associated to all end-points.

\insertplot{700}{195}
{
\ins{80pt}{100pt}{$w_6$}
\ins{140pt}{160pt}{$w_1$}
\ins{125pt}{100pt}{$w_{2}$}
\ins{100pt}{140pt}{$w_3$}
\ins{115pt}{80pt}{$w_4$}
\ins{90pt}{110pt}{$w_5$}
\ins{50pt}{80pt}{$w_7$}
\ins{40pt}{65pt}{$w_8$}
\ins{90pt}{35pt}{$w_9$}
\ins{10pt}{35pt}{$w_{10}$}
\ins{10pt}{65pt}{$w_{11}$}
\ins{40pt}{115pt}{$w_{12}$}
}
{fig61}
{\label{h2} A tree $\hat T_v$
} {0}

The vertices $w$ of $\hat T$
are then only non trivial vertices or trivial vertices with external lines $\r,\e$; all the end-points have associated an external line. 
In step 1 we consider end-points $w_a$ immediately followed by vertices $w_b$ with external lines (in the figure  
$w_4,w_{10}$
).
We can distinguish two cases. 
We call $x'_{w_a}$ and $x'_{w_b}$ the coordinate of
the external fields associated to $w_a$ and $w_b$.  If
$x'_{w_{a}}\not= x'_{w_{b}}$ we consider the vertices in the path $c_{w_a,w_b}$
in $\hat T_v$, whose number $|c_{w_a,w_b}|$ is such that
$M_{w_a,w_b}\ge A\bar x^{-\t} \g^{-h_{v'}/\t}$.
If $x'_{w_a}= x'_{w_{b}}$ we can replace the $\psi$ field in $w_b$ with a $D$ field
\be
\psi^{\e(\le h_{v}-1)}_{\r,\xx'_{w_b}} \psi^{\e(\le h_{v}-1)}_{\r,\xx'_{w_{a}}}=
\psi^{\e(\le h_{v}-1)}_{\r,\xx'_{w_b}}(
\psi^{\e(\le h_{v}-1)}_{\r,\xx'_{w_{a}}}-\psi^{\e(\le h_{v}-1)}_{\r,\xx'_{w_b}})\label{vnn}
\ee
We consider now another tree obtained canceling the end-points $w_a$
and the resulting subtrees with no external lines and we proceed in a similar way, unless the tree has no end-points followed by vertices with external lines. 
In step 2
we consider in the resulting tree  
couple of endpoints followed by the same non trivial vertex (in the picture $w_1,w_2$); we call them $w_a,w_b$ and we proceed exactly as above distinguishing the two cases. We then cancel such end-points 
and the subtrees not containing external lines, so that the end-points are associated to external lines;
we consider end-points followed by non trivial vertices with no external lines, and we proceed in a similar way. The resulting tree 
has again end-points with external lines followed by vertices with external lines (in the picture $w_5$), and we proceed as in step 1 before.
We continue in this way so that at the end all except at most one vertex with external lines are considered.
Note that by construction the different paths  $c_{w_a,w_b}$ do not overlap; 
for instance in Fig.5 the paths can be, if the corresponding coordinates are different, 
$c_{w_{10},w_{11}}$, $c_{w_4,w_5}$, $c_{w_1,w_2}$, $c_{w_5,w_{6}}$, $c_{w_6,w_{7}}$,
$c_{w_{7},w_{12}}, c_{w_9,w_{11}}$. This is an important point, as we will get a gain using the $\e$ factors associated
to each of the vertices in the path (therefore, if the path would be overlapping we would not get a gain proportional to the number of external fields).
Similarly the difference of fields involve couples of points which are not overlapping.

We consider now the vertices $v$ in $\t$ with depth increased by 1, and again we consider the corresponding tree $\hat T_v$.
The only differences with respect to the previous case is that some of the external fields 
can be $D$ fields produced by the previous step. We want to avoid that difference of fields of order greater than $1$ appear. Therefore
if the vertex is a resonant vertex and one of the external line is a difference, then 
we consider the effect of the $\RR$ operation as two separate terms.
If it has more than 6 fields, than if one of the external field is a difference field then we do not peform the operation in  the r.h.s. of \pref{vnn}. We proceed than in this way until the vertex with highest depth is reached.
At the end of this procedure, the external field of each vertex can be $\psi$ or $D$ fields (no more than a single difference of fields can be produced). Now we write each propagator involving a $D$ field 
 as in the  l.h.s. of
\pref{df11}.
We start now from the vertex $v$ with largest depth
and
we decompose the factors \be (x_0-y_0)=\sum_r (x_{0,r}-x_{0,r-1})\label{ze}\ee
along the propagators of the spanning tree $\bar T_v$. There are two possibilities: 
or such difference of coordinates correspond to the difference of coordinates of a propagator 
at scale $h_v$, or to a propagator with some scale $h_{\hat v}$; in this second case we will consider separately all the 
the field differences in the vertices between $\bar v$ and $v$.

In conclusion the final result can be written in the following way
\be V^{(h)}(\t,{\bf P})=\sum_{T\in {\bf T}}\sum_{\a\in
A_{\t,T,{\bf P}}}
 \sum_x \int dx_{0,v_0} H_{\t,{\bf P},T,\a}(x,
x_{0,v_0})\prod_{f\in P_{v_0}}\partial^{q_\a(f)} \psi^{(\le
h)\s(f)}_{\hat \xx_\a(f)}\ee
where $A_{\t,T,{\bf P}}$ is set of indices allowing to distinguish
the different terms produced by the $\RR$ operation, by the
decomposition of the zeros and by the improvements due to
anticommutativity discussed above; moreover
\bea && \sum_x \int dx_{0,v_0} H_{\t,{\bf P},T}(x, x_{0,v_0})=\\
&& \e^{n} \{\prod_{v\,\atop\hbox{\ottorm not e.p.}}{1\over
s_v!} [\prod_{v   e.p. \n }\g^{h_v}\tilde\n_{h_v}]
[\prod_{v   e.p. \z }\g^{h_v}\tilde \z_{h_v}][\prod_{v   e.p. }\g^{h_v} \tilde z_{h_v}]
[\prod_{v e.p.,\l}\tilde \l_{h_v}]
\prod_v \int
dP_{T_v}({\bf t}_v)\;{\rm det}\,\tilde  G^{h_v,T_v}({\bf t}_v)\nn\\
&&\Biggl[ \prod_{l\in T_v}
[\partial_0^{q_{0,\a}(f^-_\ell)}\partial_0^{q_{0,\a}(f^+_\ell)} ]
(\hat x_{0,l}-\hat y_{0,l})]^{b_{0,\a}(\ell)}\bar g^{(h_v)}(x''_{\ell},
\hat x_{l,0}-\hat y_{l,0})\big]_{\r^-_l,\r^+_l}\,\Bigg]
\Bigg\}\;\label{xxqq}\nn\eea
where $\hat x$ can be interpolated points and $\tilde
G^{h_v,T_v}$ is similar to $G^{h_v,T_v}$ with some  derivative applied on
the propagators
\be \tilde G^{h,T}_{ij,i'j'}=t_{ii'} \d_{x'_{ij},y'_{i'j'}}
\partial_0^{q_{0,\a}(f_{ij})}
\partial_0^{q_{0,\a}(f_{i'j'})}
\big[\tilde g^{(h)}(x_{0,ij}-y_{0,i'j'},x_{ij}
)\big]_{\r^-_{ij},\r^+_{i'j'}}\;, \label{2.48a}\ee

There is no need of a precise description of the various
contributions to the sum over $A_T$, but only need to
know some very general properties, which follows from the previous
construction.
\begin{enumerate}
\item  There is a constant $C$ such that, $\forall T\in {\bf
T}_\t$, $|A_T|\le C^n$. \item For any $\a\in A_T$, the following
inequality is satisfied
\be
\e^{n\over 4}  \Big[\prod_{f\in I_{v_0}} \g^{h_\a(f)
q_{0,\a}(f) }
\Big] \Big[\prod_{l\in T} \g^{-b_\a(l) h_\a(l)}\Big]\le \prod_{v\in H}
\g^{h_{v'}-h_v }] [\prod_{\bar v\in G, |P_v|\ge 6}\g^{-{|P_{\bar v}|}\over
6}]\label{ffg}
\ee
where $G$ is the set of non trivial vertices, or the trivial vertices in which at least two of the external fields are self-contracted.

\item $|b_0|\le 2$ and $|q_0|\le 2$
\end{enumerate}
In order to prove \pref{ffg} we note 
that the 
propagator obtained contracting a $D$ field
can be written as in the r.h.s. of \pref{df11}
so that, for any $N$, $\a\ge 0$ 
\be
|x_0-y_0|^\a |g^{(h_v)}(x_0-y_0)|\le {C_N\over 1+[\g^{h_v}|x_0-y_0|]^N} \g^{-\a h_v}\g^{-h_v}
\ee 
and
\be
|\partial^\a g^{(h_{v'})}(x_0-y_0)|\le {C_M\over 1+[\g^{h_v}|x_0-y_0|]^M} \g^{\a h_{v'}}
\ee
Therefore, with respect to the bounds in which there are no $D$ fields, one has an extra factor in the bound
\be
\Big[\prod_{f\in I_{v_0}} \g^{h_\a(f)
q_{0,\a}(f) }
\Big] \Big[\prod_{l\in T} \g^{-b_\a(l) h_\a(l)}\Big]\le \prod_v \g^{\a_v (h_{v'}-h_v)}
\ee
 where $\a_v$ is the number of $D$ fields 
external to $v$. By the construction discussed above, $\a_v\ge 1$ if $v$ is a resonant vertex
$v\in H$ so that
\be
\prod_v \g^{\a_v (h_{v'}-h_v)}\le 
[\prod_{v\in H}
\g^{h_{v'}-h_v }]\prod_{v, |P_v|\ge 6} \g^{-(\a_v -1)}
\ee
If $c_{w.w'}$ are the non overlapping paths joining two vertices $w,w'$ in $\hat T_v$ with external fields with $x_{w}\not =x_{w'}$
described above in the tree $\hat T_v$, we have
\be
\e^{n\over 2}\le \prod_{v not e.p.} \e^{N_v2^{h_{v'}-1}}\le  \prod_{c_{w,w'}} \e^{|c_{w,w'}|2^{h_{v'}-1}}\label{es}
\ee
as the total number of end-points in the set of all the paths $c_{w,w'}$ is at most equal to the number of
vertices of $\hat T_v$ (which is also the number of end-points following $v$), as the paths are not overlapping. As
$x_{w}\not=x_{w'}$ then 
$|c_{w,w''}|\ge A\bar x^{-1} \g^{-h_{v'}/\t}$ and, if $\g^{1\over\t}/2\equiv \g^\h>1$, for any vertex $\bar v\in G$ and $\e$ small enough 
\be
\e^{|c_{w,w'}|2^{h_{v'}}}\le \e^{B  \g^{-h_{v'}/\t}  2^{h_{v'}}}\le 
\g^{-1}\label{h}
\ee
In conclusion, for each vertex $v$ with $|P_v|\ge 6$
we have a factor $\g^{-1}$ (produced by the $D$ fields or by \pref{h}) proportional to $|P_v|$
so that, for a suitable constant $C$  
\be
\e^{n\over 2}\prod_{v\in G, |P_v|\ge 6} \g^{-(\a_v-1) }\le 
C^n 
[\prod_{v, |P_v|\ge 6}\g^{-{|P_{ v}|}\over
8}]
\ee
Finally the fact that $|b_0|\le 2$ and $|q_0|\le 2$ follows from the discussion below \pref{ze}.

\subsection{Bounds}

In this section we get a bound for the kernels of the effective potential defined in \pref{lau}.

\begin{lemma} If $v_h=(\tilde \l_h,\tilde \n_h,\tilde z_h,\hat \z_h)\equiv (\tilde\l_h,\tilde \a_h)$ then
\bea
&&{1\over L\b}\sum_{\t\in\TT_{h,n_\l,n_\a}}  \sum_{{\bf P}  \in{\cal P}_\t}
\sum_{T\in {\bf T}}\sum_{\a\in
A_{\t,T,{\bf P}}}
 \sum_x \int dx_{0,v_0} |H_{\t,{\bf P},T,\a}(x,
x_{0,v_0})|\le\nn\\
&& C^n |\log\e|^{n}\g^h \e^{n\over 2}|h|^n (\g^{-h}|\sup_k |\tilde\l_k||)^{n_\l} 
(\sup_k |\tilde\a_k||)^{n_a}\label{bra1}
\eea
where $C$ is a suitable constant and $n_\l,n_\a$ is the number of end-points of type $\l,\a$.
%
\end{lemma}
\vskip.3cm {\it Proof} The matrix $\tilde G^{h,T}_{ij,i'j'}$ can
be written as

\be \tilde G^{h,T}_{ij,i'j'}= \Big({\bf v}_{x_{ij}}\otimes {\bf
u}_{i}\otimes A(x_{0, ij}-,x_{ij})\;,\ {\bf v}_{y'_{i',j'}}\otimes
{\bf u}_{i'} \otimes B(y_{0,i'j'}-,x_{ij})\Big) \label{as} \;,\ee
where ${\bf v}\in \RRR^{L}$ are unit vectors such that $({\bf v}_i,{\bf v}_j)=\d_{ij}$,
${\bf u}\in \RRR^{s}$ are unit vectors $(u_i,u_{i})=t_{ii'}$, and $A,B$
are vectors in the Hilbert space with scalar product \be (A,B)=\int dz_0 A(x_0-z_0,x')B^*(z_0-y_0,x')\ee
given by
\bea &&A(x_0-z_0,x')={1\over\b}\sum_{k_0}
e^{-i k_0 (x_0-z_0)}\sqrt{f_h(k_0,x')}\;\openone\;,\nn\\
&&B(y_0-z_0,x')={1\over \b}\sum_{k_0}
e^{-ik_0 (y_0-z_0)}   \sqrt{f_h(k_0,y')}\Big[A_h(k_0,x')\Big]^{-1}\;.
\label{2.48b}\eea
with $A_h$ defined in (67).
Therefore
\be |{\rm det}\tilde  G^{h_v,T_v}({\bf t}_v)| \le\bar C^n\ee

We write the factor $\e^n$ in \pref{xxqq} as $\e^{n\over
2}\e^{n\over 4}\e^{n\over 4}$; we write $\e^{n\over 4}$ using
Lemma 3.3 while the other factor is used in \pref{ffg}; therefore

\bea &&{1\over L\b}\sum_x \int dx_{0,v_0} |H_{\t,{\bf P},T,\a}(x,
x_{0,v_0})|\le [\prod_v {1\over s_v!}] [ \prod_{v\in L} \e^{A
\bar x^{-1} \g^{-h_{v'}\over \t}2^{h_{v'}}}
][\prod_{v\in H} \g^{h_{v'}-h_v }]\prod_{v}\g^{-\a |P_v|  }\nn\\
&&[\prod_{v, S_v>1} \g^{-h_v(S_v-1)}]
[\prod_{v   e.p. \a }\g^{h_v}|\tilde \a_{h_v}|][\prod_{v e.p. \l} |\tilde \l_{h_v}|] \e^{ n\over 2}
\eea
where $L$ is the set of non resonant clusters.  We use that
\be
[\prod_{v\in H} \g^{h_{v'}-h_v }]=[\prod_{v\in H, S_v>1} \g^{h_{\bar v'}-h_v }]
\ee
where $\bar v'$ is the first non trivial vertex following the non trivial vertex $v$ (note that all the vertices between $v$ and $\bar v'$ are resonant).
Note that
\be
[\prod_{v, S_v>1} \g^{-h_v(S_v-1)}]
[\prod_{v\in H, S_v>1}\g^{h_{\bar v'}-h_v}]\le  \g^{h}
[\prod_{v, S_v>1} \g^{-h_v S_v}]
[\prod_{v\in H, S_v>1}\g^{h_{\bar v'}}]
\ee
where we have used that no $\RR$ operation acts on $v_0$
Moreover by using \pref{bra}
\be
\prod_{v, S_v>1} \g^{-h_v S_v}=[\prod_{v, S_v>1} \g^{-h_v S^H_v}][\prod_{v, S_v>1} \g^{-h_v S^L_v}][\prod_{v, S_v>1} \g^{-h_v S^2_v}]
\ee
Note that
\be
[\prod_{v, S_v>1} \g^{-h_v S^H_v}]
[\prod_{v\in H, S_v>1}\g^{h_{\bar v'}}]= 1
\ee
Moreover by \pref{zzaa}
\be
[\prod_{v, S_v>1} \g^{-h_v S^L_v}]
[
\prod_{v\in L}
\e^{A \bar x^{-1} \g^{-h_{v'}\over \t}2^{h_{v'}}}]\le 
\bar C^n\prod_{v, S_{ v}>1}  \g^{-h_v S^L_v}
 \g^{3 S^L_{v} h_{ v}}\le \bar C^n
\label{zzaa} \ee
following from the fact that, 
as $\g^{1\over\t}/2\equiv \g^\h>1$, for any $N$
\be
\e^{A \bar x^{-1} \g^{-h\over \t}2^{h}}=e^{-|\log\e|
A \bar
x^{-\t}\g^{-\h h}}\le \g^{N \h h} [{N\over |\log\e| A \bar
x^{-\t}]^N e^{-N}}
\ee
as  $e^{-\a x} x^N\le [{N\over \a}]^N e^{N}$. Therefore, by choosing $N=3$ we get
$\prod_{v\in L} \e^{A \bar
x^{-\t} \g^{-h_{v'}\over \t}2^{h_{v'}}} \le \bar  C^n\prod_{v, S_v>1} \g^{3 S^L_{ v} h_{v}}$
where $\bar C= [{3\over \log\e A \bar x^{-\t]}}]^3 e^{3}]$ and we have used that the number of non trivial vertices is smaller of the number of end-points $n$.
%
%
Finally
\be
[\prod_{v not e.p} \g^{-h_v S^2_v}][\prod_{v   e.p. \a}\g^{h_v}|\a_{h_v}|]
[\prod_{v e.p.,\l} |\tilde \l_{h_v}|] \le C^n
[\prod_{v   e.p. \a}|\tilde \a_{h_v}|]
[\prod_v |\g^{-h_v}\tilde \l_{h_v}|
\ee
Therefore
we get the bound
 \be
{1\over L\b}
 \int dx_{v_0}| W_{\t,\PP, T}(\xx_{v_0})|\le C^n \e^{n\over 2}\g^h 
 [\prod_{v e.p.}
|\tilde\a_{h_v}|][\prod_{v e.p.} \g^{-h_v}|\tilde \l_{h_v}|\prod_{v not e.p.}\g^{-\a |P_v|  }
\ee
The sum over ${\bf  P}$ is done as in \pref{B.18c}; the sum over $\a$ is over $C^n$ terms.
The sum over the trees $\t$ is done performing the sum of unlabelled
trees and the sum over scales. The latter can be bounded by $|h|^m$, where $m$ is the number
of non trivial vertices, which is $\le C^n$; indeed given the
unlabeled tree, the scales of the trivial vertices and of the
end-points are determined once that the scales of the non trivial
vertices are determined; the former is bounded by $C^n$ so that \pref{bra1} follows.
%
%
\qed
\subsection{The flow of the effective coupling}

It is an easy consequence of Lemma 3.5 the following result
\begin{lemma} If $\g^{\bar h}\ge \e^{2\bar x}$
then there exists an $\e_0$ and a choice $\n$ such that for $\e\le \e_0$ and $|\l|\le \e^{2\bar x+2}$ then there exists a suitable constant $C_1$ such that, for any $k\ge \bar h$
\be
|\tilde\l_h|\le |\tilde\l|C_1\quad |\hat\a_h|\le C_1\label{xa}
\ee
\end{lemma}
{\it Proof} We proceed by induction. 
The flow equation for $\n_k$ is 
\be
\tilde\n_{k-1}=\g \tilde\n_h+\g^{-k}\int dx_0  H^{(k)}_{2,\r\r}(0,x_0,0)
\ee
with $\tilde\n_2=\tilde\n$.  By iteration we get 
\be
\tilde\n_k=\g^{-k+1}(\tilde\n+\sum_{k'\ge k}\int dx_0  H^{(k')}_{2,\r\r}(0,x_0,0))
\ee
and by properly choosing $\tilde\n$ so that $\n_{\bar h}=0$ we get
\be
\tilde\n_k=-\g^{-k+1}\sum_{\bar h\le k'\le k}\int dx_0  H^{(k')}_{2,\r\r}(0,x_0,0))\label{lau1}
\ee
and one can show by a fixed point argument, the existence of a bounded sequence of $\tilde\n_k$ verifying
\pref{lau1} (the proof is identical to the one  
\S A2.6 of \cite{GM}).
Regarding the flow of $\tilde\z_h$
assume that \pref{xa} is true for $k\ge h$. The flow equation for $\tilde\z_h$ is
\be
\tilde\z_h=\sum_{k\ge h}\int dx_0  \partial H^{(k)}_{2}
\ee
%
Using lemma  3.4 and the fact that the derivative cancels a factor $\g^h$ we get
for $\e$ small enough
\be
|\tilde\z_h|\le\sum _{n=2}^\io\sum_{k\ge h}
 C^n C_1^n\e^{n\over 4}|h|^n 
\le |h| C_2 (C C_1 |h| \e^{1\over 4})\le  C_1
\ee
where we use that $|h|\e^{1\over 4}\le\e^{1\over 8}$ and $\g^{-k}|\tilde \l|\le\e$.
\vskip.3cm
Similarly
\bea
&&|\tilde\l_h|\le|\l_0|+\sum _{n=2}^\io\sum_{n_\l\ge 1}\sum_{k\ge h}
C^n C_1^n\g^k \e^{n\over 4}|h|^n (\g^{-k} |\tilde\l_k|)^{n_\l}\le\nn\\
&&\sum_{n=2}^\io |h|^{n+1}\e^{n\over 4}C^n C_1^n
\sum _{n_\l=1}^\io  |\l| (\g^{-h}|\tilde\l_k||)^{n_\l-1}
\le |\l| C_1
\eea
\qed
\vskip.3cm
The above lemma says the the flow is bounded up to 
a scale  $\g^h\ge \e^{2\bar x}$. In order to integrate the smaller scales one has to use the mass term.
Note indeed that if there exists two constants such that \be c_1\e^{2\bar x}\le \s_h\le c_2\e^{2\bar x}\label{fon}\ee then
there exists a scale $h^*$ with $\g^{h^*}=O(\e^{2\bar x})$ defined as 
the minimal $h$ such that 
$\g^{k}\ge \s_{k}$.
For any integer $M$  and a
suitable constant $C_M$
\be
|\bar g^{(\le h^*)}(x_0-y_0,x)| \le {C_M\over 1+(\g^{h^*}|x_0-y_0|)^M}\label{bo}
\ee
Indeed the denominator of the propagator is $\ge c \g^{h^*}$,
so that the above bounds follow using integration by parts.
The bound \pref{bo} says that the propagator corresponding to all the scales $\le h^*$
verifies the same bound of the single scale propagator; therefore we can bound the scale $\le h^*$
in a single step, and, by lemma 3.4 and 3.5 (for $h>h^*=\log\e^{2\bar x}$), 
convergence follows. It remains to prove \pref{fon} and that $\bar\z_h=O(\e^{2\bar x})$.
We can write  (similar expressions hold for $\bar z_h$)
\be
\s_h=\sum_{k\ge h}\int dx_0  H^{(k)}_{2,\r-\r}(0,x_0,0)
\ee
and
\be
H^{(h)}_2=H^{(a)(h)}_{2,\r,-\r}+H^{(b)(h)}_{2,\r,-\r}
\ee
where $H^{(a)(h)}_{2,\r,-\r}$ is the sum over trees with
$n\le 8\bar x$ and $H^{(b)(h)}_{2,\r,-\r}$
is the sum over trees with $n\ge  8\bar x+1$. The bound for 
$H^{(b)(h)}_{2,\r,-\r}$ from lemma 3.4 is $\le C \e^{2\bar x+{1\over 4}}$.
Regarding $H^{(a)(h)}_{2,\r,-\r}$ we again distinguish between trees with at least a $\l,\l_h$ end-point and the rest; the former is bounded by $C \g^h|\g^{-h}\tilde\l|\le C \e^{2\bar x+1}$.
Regarding the latter, it can be represented in terms of chain graphs with $\e,\n,\z$ end-points
;  if $x'_\ell$ is the coordinate of any internal propagator with scale $h$ and $x'$ is the external coordinate,
$x\not =x'$ , $a$ is a constant
\be
a\g^h\ge ||\o x'||_1+||\o x'_{\ell}||_1\ge  ||\o x'-\o x'_{\ell}||\ge C_0 |x'-x_\ell|^{-\t}\ge C_0 |(8\bar x)^2|^{-\t}\label{ff}
\ee
Moreover the graphs are $O(\e^{k})$ with $k\ge 2\bar x$ (the sum of coordinates of internal vertices is $2\bar x$, and the difference of coordinate of the terms attached to $\e$ vertices is $\pm 1$, and to $\z$ is $\pm 2\bar x$, but they are $O(\e^{2\bar x}))$. There is only one term contributing to lowest order;
its value is $\e^{2\bar x}a$ with
\be
a={\chi_{\ge h}\over \phi_{-\bar x+1}-\phi_{\bar x}}{\chi_{\ge h}\over \phi_{-\bar x+2}-\phi_{\bar x}}.....{\chi_{\ge h}\over \phi_{\bar x-1}-\phi_{\bar x}}
\ee
where $\chi_{\ge h}$ is the cut-off function $\chi_{u,v}+\sum_{\r=\pm} \sum_{k=h}^0 f^{(k)}(\o (x-\r\bar x))$. The terms proportional to $\e^k$, with $2\bar x+1\le k\le 8\bar x$ 
have at most $8\bar x$ propagators bounded by \pref{ff}; therefore
\be
\s_h=\e^{2\bar x}(a+O(\e \bar x!^\a)+O(\e^{2\bar x+{1\over 4}}))\label{xa1}
\ee
Therefore, for $\e\le O(\bar x!^{-\a})$ 
then \pref{fon} follows.

\subsection{The 2-point function}

We have finally to get a bound for the two-point function. 
First of all, we note that 
Lemma 3.4 and Lemma 3.5 immediately imply  a bound 
for the kernel of the effective potential with two external lines, with coordinate $x$ 
and $y$. Indeed 
in the trees $\t\in\TT_{h,n}$ with $n$ end-points
contributing to 
$W^{(h)}_2$ there is necessarily a path $c_{w_1,w_2}$ in $\hat T_v$ connecting the points $w_1$, with $\xx_{w_1}=\xx$
and $w_2$ with $\xx_{w_2}=\yy$ such that by \pref{fa} $|x-y|\le 8 |c_{w_1.w_2}\bar x|$;
moreover $|c_{w_1,w_2}|\le n$ so that $n\ge {1\over 8\bar x}|x-y|$.
Therefore 
no tree $\t$ with 
 $n< {1\over 8\bar x}|x-y|$ contribute to a kernel of the effective potential with external lines with coordinate $x$ and $y$; therefore by Lemma 3.4 and Lemma 3.5 we get, for $h\ge h^*$
\be
{1\over \b}\int dx_0 |W^{(h)}_2(\xx,\yy)|\le
\sum_{n\ge {1\over 8\bar x}|x-y| }
C^n |h|^n\g^h
|\log\e|^{n} \e^{n\over 2}|h|^n\le  C \g^{h}\e^{\a |x-y|}\label{ss}\ee
with suitable $\a$ and $C$.

In order to bound the 2-point function we have to consider the multiscale integrtion with $\phi\not=0$;
we get
\be
S_2(\xx,\yy)=\sum_{h=h^*}^1 S_{2,h}(\xx,\yy)
\ee
and $S_{2,h}(\xx,\yy)$ are expressed in terms of a tree expansion similar
to the one for $W_2^{(h)}$, where the only difference is that two external fields are replaced by propagators
$g^{(k)}(x';x_0-z_0)$ and $g^{(l)}(y';x_0-z_0)$; 
therefore 
$S_{2,h}(\xx,\yy)$ (at $\xx,\yy$ fixed) verifies a bound similar to \pref{ss} with an extra
extra factor $C_N {\g^{-h}\over 1+\g^{N h}|x_0-y_0|^{N}}$
for any $N$, that is 
\be
|S_{2,h}(\xx,\yy)|\le \e^{\a |x-y|}{C_N \over 1+\g^{N h}|x_0-y_0|^{N}}
\ee
In conclusion, by \pref{xa1}, for any $N$
\be
|S_2(\xx,\yy)|\le \sum_{h=h^*}^0\e^{\a |x-y|}{C_N \over 1+\g^{N h}|x_0-y_0|^{N}}
\le \tilde C _N{e^{-{\a\over 2}|\log \e||x-y|}\over 1+[\s_{h^*}|x_0-y_0|]^N}
\ee
so that \pref{xx1} is proved.


\begin{thebibliography}{99999}

\bibitem{A} P. W. Anderson Phys. Rev. 109, 1492 (1958)

\bibitem{loc0}
J. Froehlich and T. Spencer, 
, Comm. Math. Phys. 88 (1983), 151.

\bibitem{loc1}
M. Aizenman and S. Molchanov, 
Comm. Math. Phys. 157 (1993), 245

\bibitem{AA}
S. Aubry and G. Andre, Ann. Israel Phys. Soc 3, 1 (1980).


\bibitem{DS} Dinaburg, E, Sinai Y Funct analysisi and its app. 9, 279 (1975)


\bibitem{E} L.H. Eliasson Comm. Math. Phys 146, 447 (1992)

\bibitem{BLT} J. Bellissard, R. Lima, and D. Testard, Comm.
Math. Phys. 88, 207 (1983).

\bibitem{S} Sinai, Ya., J. Stat. Phys. 46, 861 (1987)

\bibitem{FS}
Froehlich, J., Spencer, T. Wittvwer Comm. Math. Phys. 88 (1983), 151–189.

\bibitem{Si} Sinai, J. Stat. Phys. 46 (1987), 861–909.

\bibitem{Av}  A. Avila, S. Jitomirskaya Ann. of. Math. 170
 303-342 (2009)


\bibitem{FA} L.Fleishmann, P.W. Anderson Phys. Rev B 21, 2366 (1980)


\bibitem{GT}
T. Giamarchi, T., and H.J. Schulz, Europhys. Lett. 3 (1987), 1287.


\bibitem{GGG}
Gornyi, I., Mirlin, A., and Polyakov, D. 
Phys. Rev. Lett. 95 (2005), 206603.


\bibitem{VMG}J. Vidal, D. Mouhanna, T. Giamarchi  Phys, Rev. Lett. 83, 3908 (1999)



\bibitem{B} D.M. Basko, IL. Alteiner , B. L. Altshuler Ann. of Physics
Ann. Phys. (N. Y). 321, 1126 (2006)

\bibitem{PH} Pal, A., and Huse, Phys. Rev. B 82, (2010),
174411.


\bibitem{NH} R. Nandkshore, D. Huse . arXiv:1404.0686



\bibitem{H4}
S. Iyer, V. Oganesyan, G. Refael, D. A. Huse Phys. Rev. B 87, 134202 (2013)







\bibitem{loc2}
Fauser, M., and Warzel, S. 
arXiv:1402.5832.


\bibitem{loc4}J. Imbrie. 
arxiv1403.7837

\bibitem{loc4a} T. Spencer. Talk at a conference in Rome 2012

\bibitem{loc3} De Roeck, W., and Huveneers, F. 
arXiv:1308.6263 

\bibitem{G} G. Gallavotti. Comm.Math. Phys 164, 1, 145-156 (1994)

\bibitem{GM1} G.Gentile, V.Mastropietro. Rev. Math. Phys. 8, 3, 393-444 (1996).

\bibitem{BGM} G.Benfatto, G. Gentile, V.Mastropietro. J. Stat. Phys. 89, 655-708 (1997)


\bibitem{GM} G. Gentile, V.Mastropietro. Comm. Math.Phys. 215, 69-103   (2000)

\bibitem{M} V.Mastropietro. Comm. Math. Phys. 201, 81 (1999)





\bibitem{BFM}  G.Benfatto, P.Falco V.Mastropietro. Comm. Math.Phys. 330, 153-215 (2014)

\end{thebibliography}
\end{document}